\documentclass[twocolumn,final,prb,superscriptaddress,eqsecnum,10pt
amssymb,amsmath,amsfonts,showpacs,floatfix,citeautoscript]{revtex4}
\usepackage{graphicx,epstopdf}
\usepackage{subfigure}
\usepackage{mathrsfs}
\usepackage{bm}
\usepackage[latin1]{inputenc}
\usepackage{wasysym}    
\usepackage{color}

\newcommand{\beq}{\begin{equation}}
\newcommand{\eeq}{\end{equation}}
\newcommand{\nn}{\nonumber}

\newcommand{\be}{\beta}

\definecolor{mygreen}{rgb}{0,0,0}

\def\be{\begin{equation}}
\def\ee{\end{equation}}
\def\bc{\begin{center}}
\def\ec{\end{center}}
\newcommand{\bea}{\begin{eqnarray}}
\newcommand{\eea}{\end{eqnarray}}

\def\vec{\mathbf}

\newcommand{\dagga}{{\phantom{\dagger}}}

\usepackage[colorlinks,bookmarks=false,citecolor=blue,linkcolor=red,urlcolor=blue]{hyperref}
\begin{document}
\title{Phase diagram of hard-core bosons on clean and disordered 2-leg ladders\\
Mott insulator -- Luttinger liquid -- Bose glass}
\author{Fran\c cois Cr\'epin}
\email{crepin@lps.u-psud.fr}
\affiliation{Laboratoire de Physique des Solides, Universit\'e Paris-Sud, UMR-8502 CNRS, 91405 Orsay, France}
\author{Nicolas Laflorencie}
\email{laflo@irsamc.ups-tlse.fr}
\affiliation{Laboratoire de Physique des Solides, Universit\'e Paris-Sud, UMR-8502 CNRS, 91405 Orsay, France}
\author{Guillaume Roux}
\email{guillaume.roux@u-psud.fr}
\affiliation{Laboratoire de Physique Th\'eorique et Mod\`eles Statistiques, Universit\'e Paris-Sud, UMR-8626 CNRS, 91405 Orsay, France}
\author{Pascal Simon}
\email{simon@lps.u-psud.fr}
\affiliation{Laboratoire de Physique des Solides, Universit\'e Paris-Sud, UMR-8502 CNRS, 91405 Orsay, France}
\date{\today}
\begin{abstract}
One dimensional free-fermions and hard-core bosons are often considered to be equivalent. Indeed, when restricted to nearest-neighbor hopping on a chain the particles cannot exchange themselves, and therefore hardly experience their own statistics. Apart from the off-diagonal correlations which depends on the so-called Jordan-Wigner string, real-space observables are similar for free-fermions and hard-core bosons on a chain. Interestingly, by coupling only two chains, thus forming a two-leg ladder, particle exchange becomes allowed, and leads to a totally different physics between free-fermions and hard-core bosons. Using a combination of analytical (strong coupling, field theory, renormalization group) and numerical (quantum Monte Carlo, density-matrix renormalization group) approaches, we study the apparently simple but non-trivial model of hard-core bosons hopping in a two-leg ladder geometry. At half-filling, while a band insulator appears for fermions at large interchain hopping $t_{\perp}>2t$ only, a Mott gap opens up for bosons as soon as $t_{\perp}\neq 0$ through a Kosterlitz-Thouless transition. Away from half-filling, the situation is even more interesting since a gapless Luttinger liquid mode emerges in the symmetric sector with a non-trivial filling-dependent Luttinger parameter $1/2\le K_s\le 1$. Consequences for experiments in cold atoms, spin ladders in a magnetic field, as well as disorder effects are discussed. In particular, a quantum phase transition is expected at finite disorder strength between a 1D superfluid and an insulating Bose glass phase.
\end{abstract}
\pacs{05.30.Jp,75.10.Jm,05.30.Rt}
\maketitle
\section{Introduction}
\label{sec:intro}
Low-dimensional bosonic systems have attracted an increasing interest during the last few decades~\cite{Fisher89,Nozieres-book,Greiner02,Leggett-book,Bloch08}. The competition between various ingredients such as interactions, quantum fluctuations, geometrical constraints (low dimensionality/coordinance, frustration), possibly disorder, may lead to a large variety of interesting states of matter, as for instance the enigmatic supersolid phase of $^4$He under pressure~\cite{Balibar10}. Perhaps the most extreme situation arises for one-dimensional (1D) systems where interactions are known to induce dramatic effects~\cite{Giamarchi-book,Cazalilla11}. While 1D systems have long been considered as purely abstract objects reserved for theoretical studies, they have now become experimentally accessible.
In particular, due to a very intense activity, several systems are now available in the lab to achieve 1D bosonic physics in various contexts: Josephson junction arrays~\cite{Chow98,Fazio01,Refael07}, superfluid $^4$He in porous media~\cite{Gordillo00,Savard09,Taniguchi10,DelMaestro10,DelMaestro11}, cold atoms~\cite{Jaksch98,Greiner02,Paredes04,Roati08,Billy08,Chen11}, quantum antiferromagnets~\cite{Chaboussant97,Mila98,Hikihara2001,Affleck05,Garlea07,Klanjsek08,Ruegg08,Giamarchi08,Hong10,Bouillot11}. Furthermore, several theoretical studies have recently shown that 1D physics is still (80 years after Bethe's seminal work~\cite{Bethe31}) a very active playground for exploring novel and exotic phases of matter~\cite{Sheng08,Sheng09,Block11,Chen11}.

As in most condensed matter problems, only a few competing terms dictate the physics of 1D bosonic systems. For instance, when kinetic processes dominate over (repulsive) interactions, a delocalized superfluid (SF) phase is favored. Note that despite the absence of a true Bose-condensed phase in 1D systems~\cite{Hohenberg67}, superfluidity is still possible at zero temperature since it only requires off-diagonal quasi long-range order, the resulting phase being a Luttinger liquid (LL). Conversely, when strong repulsion dominates, a Mott insulator (MI) is expected, either breaking or not lattice symmetries, depending on the commensurability of the particle filling. 
\begin{figure}
\includegraphics[width=6cm,clip]{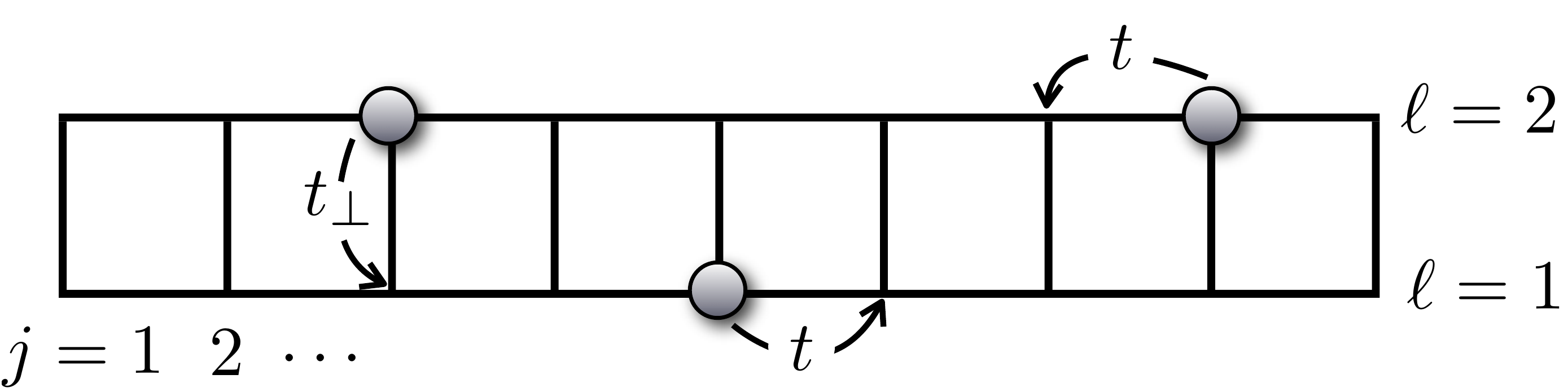}
\caption{\label{fig:schem}(Color online) Schematic picture for the two-leg ladder system.}
\end{figure}

An interesting case in 1D is the hard-core boson limit (also known as the Tonk-Girardeau gas, as achieved in cold atom experiments~\cite{Paredes04}) which, on a lattice, corresponds to infinite on-site repulsion. In such a limit, and in  the absence of density-density interactions between neighboring sites, an exact mapping between fermions and hard-core bosons~\cite{Girardeau60} implies that bosons behave "almost" like free-fermions, at least regarding their real space properties, where a ``real space Pauli principle'' holds. However, when the lattice geometry allows for particle exchange,
the bosonic statistics may play a major role, leading to qualitatively different phase diagrams. It is quite clear in two dimensions~\cite{Fradkin89,Bernardet02}, but for quasi-1D geometries like ladders one may wonder whether a minimal model made of two coupled chains would allow the bosonic statistics to emerge and lead to qualitatively different physics as compared to free-fermions.

\begin{figure}
\includegraphics[width=\columnwidth,clip]{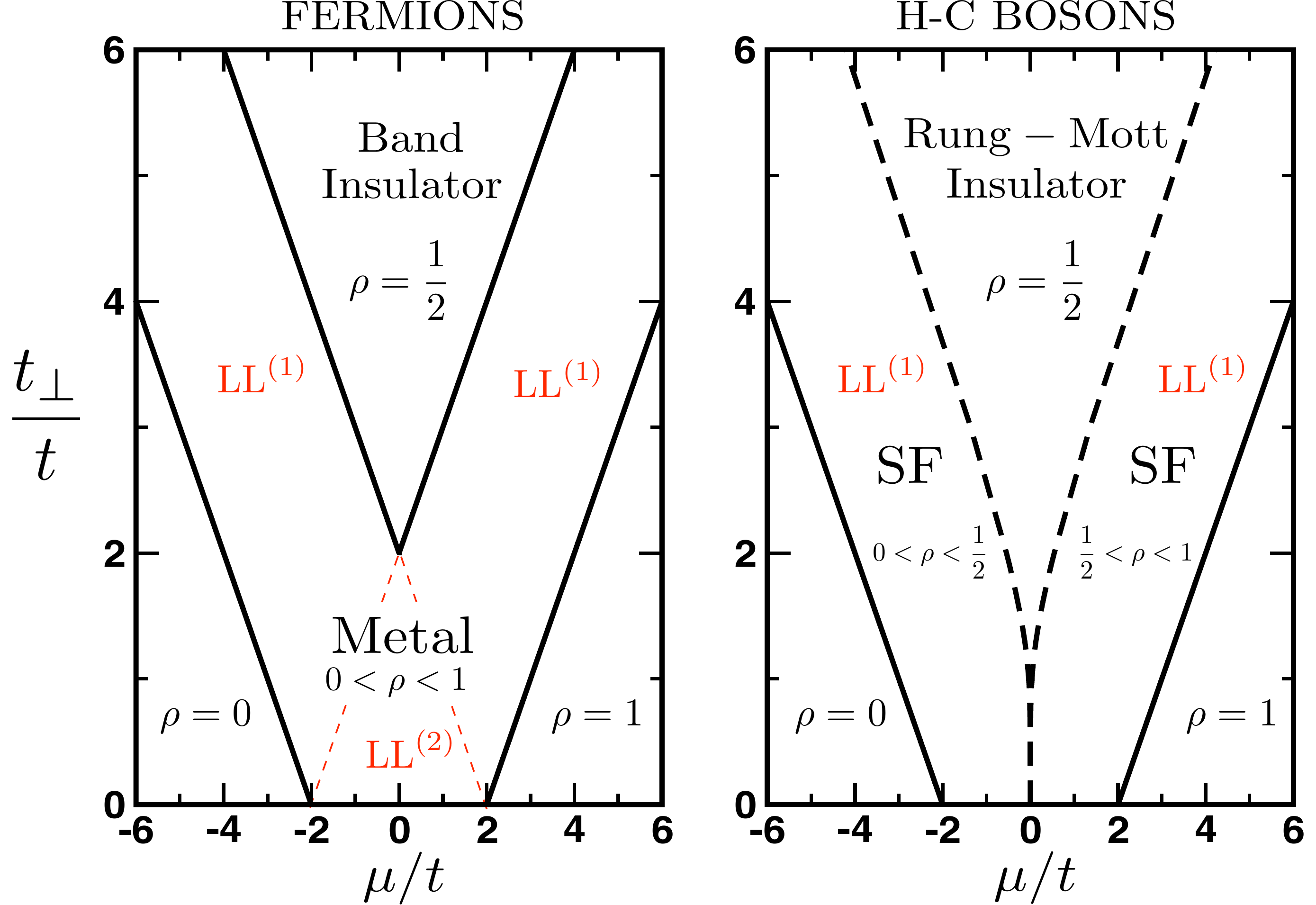}
\caption{\label{fig:PHDG}(Color online) Phase diagram of the two-leg ladder model Eq.~(\ref{eq:model}) for free-fermions (left) and hard-core bosons (right). The metallic phase for fermions comprises either one LL$^{\rm{(1)}}$ or two bands LL$^{\rm{(2)}}$, whereas the bosonic SF is a single mode (symmetric, see the text) Luttinger liquid LL$^{\rm{(1)}}$. For hard-core bosons, the rung-Mott insulating phase extends down to $t_\perp=0$.}
\end{figure}

In this paper, we aim at exploring one of the simplest quasi-1D model where free-fermions and hard-core bosons display qualitatively distinct phase diagrams. The model, defined on a two-leg ladder (depicted in Fig.~\ref{fig:schem}), is governed by the following Hamiltonian
\bea
{\cal{H}}&=&
-t\sum_{j,\ell=1,2}\big[b^{\dagger}_{\ell,j}b_{\ell,j+1}^{\dagga}+{\rm h.c.}\big]-\mu\sum_{j,\ell=1,2}n_{\ell,j}\nonumber\\
&&-t_{\perp}\sum_{j}\big[b^{\dagger}_{1,j}b_{2,j}^{\dagga}+{\rm h.c.}\big],
\label{eq:model}
\eea
where the sites along each leg $\ell=1,2$ are labeled by $j=1\ldots L$, $L$ being the total length of each chain. The operator $b^{\dagger}_{\ell,j}$ creates a particle on site $(\ell,j)$ and $t$, $t_\perp$ are the longitudinal and transverse hopping integrals respectively. $n_{\ell,j} = b^{\dagger}_{\ell,j}b_{\ell,j}$ is the onsite density operator and $\mu$ is the chemical potential. The filling of the ladder will be denoted by $\rho = N/(2L)$ in the following, with $N$ the total number of bosons. The main phases of \eqref{eq:model}, discussed in this work, are summarized on Fig~\ref{fig:PHDG} in the cases of free-fermions (left) and hard-core bosons (right). For non-interacting fermions, the phase diagram is trivially obtained by filling two bands: the system is a simple metallic state at all fillings $0<\rho<1$, except when the interchain hopping $t_\perp>2t$ where a band insulator appears at half-filling $\rho=1/2$. 
The situation is very different with hard-core bosons for which there is no Pauli principle when filling the states in momentum space. Indeed, as demonstrated below, the interchain hopping $t_\perp$ is shown to be a relevant perturbation at half-filling, opening up a charge gap  $\Delta_s\sim \exp(-at/t_\perp)$ via a Kosterlitz-Thouless mechanism. Therefore, in contrast with free-fermions, the bosonic insulating state at half-filling, called {\it rung-Mott insulator}, extends down to $t_\perp=0$. For incommensurate fillings, the bosonic system is a single mode LL (called LL$^{\rm{(1)}}$ by contrast with LL$^{\rm{(2)}}$ for the two-band or two-mode Luttinger liquid of free-fermions, see Fig.~\ref{fig:PHDG}) with a finite superfluid (SF) fraction and, most interestingly, a strongly renormalized Luttinger parameter $K(\rho)$ which varies continuously between $1/2$ and $1$.

In the rest of the paper, using a combination of analytical and numerical techniques, we discuss in details all these non-trivial aspects. In section~\ref{sec:1D}, after recalling the similarities between free-fermions and hard-core bosons on a chain, we present some well-known exact results for free-fermions on a ladder. In section~\ref{sec:strong_rung}, the two coupled chains problem is studied for hard-core bosons in the two analytically accessible limits: strong and weak interchain couplings, using a perturbative approach, bosonization and renormalization group (RG) techniques. We also compare the physics of this model with the well known case of quantum spin ladders. We then use exact numerical techniques, quantum Monte Carlo and density-matrix renormalization group, and confront them with analytical predictions in order to provide quantitative results for several quantities of importance: the bosonic densities (normal and SF), the charge gap, correlation functions, correlation lengths, and Luttinger parameters. In particular, section~\ref{sec:RMI} focuses on the half-filled insulating state while the LL behavior at incommensurate fillings is studied in great details in section~\ref{sec:LL} where the global phase diagram is shown. Finally, in section~\ref{sec:disc} we discuss two important issues related to experiments in cold atoms and spin ladder materials as well as the effect of disorder on this bosonic ladder. Section~\ref{sec:conc} concludes the paper.
%

\section{Free-fermions {\it vs} hard-core bosons}
\label{sec:1D}
\subsection{Quantum statistics in 1D systems}
The concept of quantum statistics stems from the indistinguishability of identical particles in many-body quantum systems. If $\Psi(\mathbf{x}_1,\mathbf{x}_2)$ is the wave-function of a system made of two indistinguishable particles, the probability of finding two particles at position $\mathbf{x}_1$ and $\mathbf{x}_2$ must be invariant under the exchange or particles, that is, $|\Psi(\mathbf{x}_1,\mathbf{x}_2)|^2 = |\Psi(\mathbf{x}_2,\mathbf{x}_1)|^2$. Therefore $\Psi(\mathbf{x}_1,\mathbf{x}_2)$ and $\Psi(\mathbf{x}_2,\mathbf{x}_1)$ are equal, up to a phase factor $e^{i\alpha}$. In three dimensions, it can be argued that a further exchange of particles is actually equivalent to the identity transformation, imposing that $\alpha$ be an integer multiple of $\pi$. Two situations arise: the wave-function is either symmetric or antisymmetric under the exchange of two particles, thus defining bosonic and fermionic statistics, respectively. Note that the present restriction does not hold in lower dimensions where so-called {\it fractional} statistics are expected to exist. However, even though we will only be interested in 1D and quasi 1D systems we will focus on bosons and fermions only. This real-space approach of quantum statistics extends to any Hilbert space in that a given many-body state -- a Fock state -- will either be totally symmetric or totally antisymmetric under the exchange of any two particles. As a consequence, fermions obey Pauli's exclusion principle, while several bosons can indeed occupy the same quantum state -- actually a macroscopic number in the phenomenon of Bose-Einstein condensation. There is one situation though, where differences between bosons and fermions are less drastic. Indeed, a 1D gas of impenetrable bosons exhibits similarities with a 1D gas of spinless fermions. The hard-core constraint mimics the exclusion principle, while the 1D character ensures that the particles cannot {\it move around} each other. Wave-functions are identical up to a {\it symmetrization} factor enforcing the quantum statistics. Therefore, quantities depending only on the modulus of the wave-function, such as density correlations, are identical, whereas {\it off-diagonal} quantities, such as the momentum distribution, are affected by statistics. These various quantities -- density and momentum distribution -- are readily encoded in the single-particle density-matrix $\langle{\psi^\dagger(x)\psi(x')}\rangle$, where $\psi^\dagger(x)$ and $\psi(x)$ are creation and annihilation operators.

\subsubsection{When fermions and hard-core bosons look alike}
In the continuum, if $\Psi_F(x_1,\ldots,x_N)$ is a solution of the Schr\"{o}dinger equation for a 1D gas of $N$ fermions, satisfying the constraint $\Psi_F(x_1,\ldots,x_N)=0$ whenever $x_i=x_j$ for all $i$ and $j$, then $\Psi_B(x_1,\ldots,x_N)= A(x_1,\ldots,x_N)\Psi_F(x_1,\ldots,x_N)$ is also a solution of the  Schr\"{o}dinger equation for a 1D gas of hard-core bosons, with $A=\prod_{i>j}\textrm{Sign}(x_i-x_j)$, an antisymmetric function which can only be equal to $\pm 1$. \cite{Girardeau60} It appears clearly that $|\Psi_F(x_1,\ldots,x_N)|^2 = |\Psi_B(x_1,\ldots,x_N)|^2$, so that any quantity depending only on the modulus of the wave function is identical for fermions and hard-core bosons. As noticed by Girardeau, \cite{Girardeau60,Girardeau65} this one to one correspondence between fermionic and bosonic wave-functions is only possible in one dimension. The hard-core constraint divides the parameter space in $N!$ disjoint regions, and $A$ changes sign discontinuously at the boundary of each regions. In two and three dimensions, the parameter space remains connected and a function equivalent to $A$ cannot be defined. On a 1D lattice the correspondence is ensured through the Jordan-Wigner transformation\cite{Jordan1928}. Hard-core bosons on a lattice have mixed commutation relations: $[b_i,b^\dagger_j]=[b_i,b_j]=[b^\dagger_i,b^\dagger_j]=0$ for $i\neq j$, $\{b_i,b^\dagger_i\} = 1$ and $(b_i)^2 = (b^\dagger_i)^2 = 0$. The Jordan-Wigner transformation maps this problem of hard-core bosons onto a problem of spinless fermions $c_l$ through:
\begin{equation}
c_l = \exp\Big[i\pi \sum_{j=1}^{l-1}b^\dagger_j b_j\Big]\,b_l\;.
\end{equation} 
The Jordan-Wigner string, although making the transformation non-local, ensures that $c_l$ and $c^\dagger_l$ satisfy anticommutation relation and are indeed fermionic operators. 
As in the continuum case, there is a one to one correspondence between the eigenstates of the Fermi Hamiltonian and the eigenstates of the Bose Hamiltonian. 

\subsubsection{When fermions and hard-core bosons look different}

\begin{figure}
\includegraphics[width=0.8\columnwidth,clip]{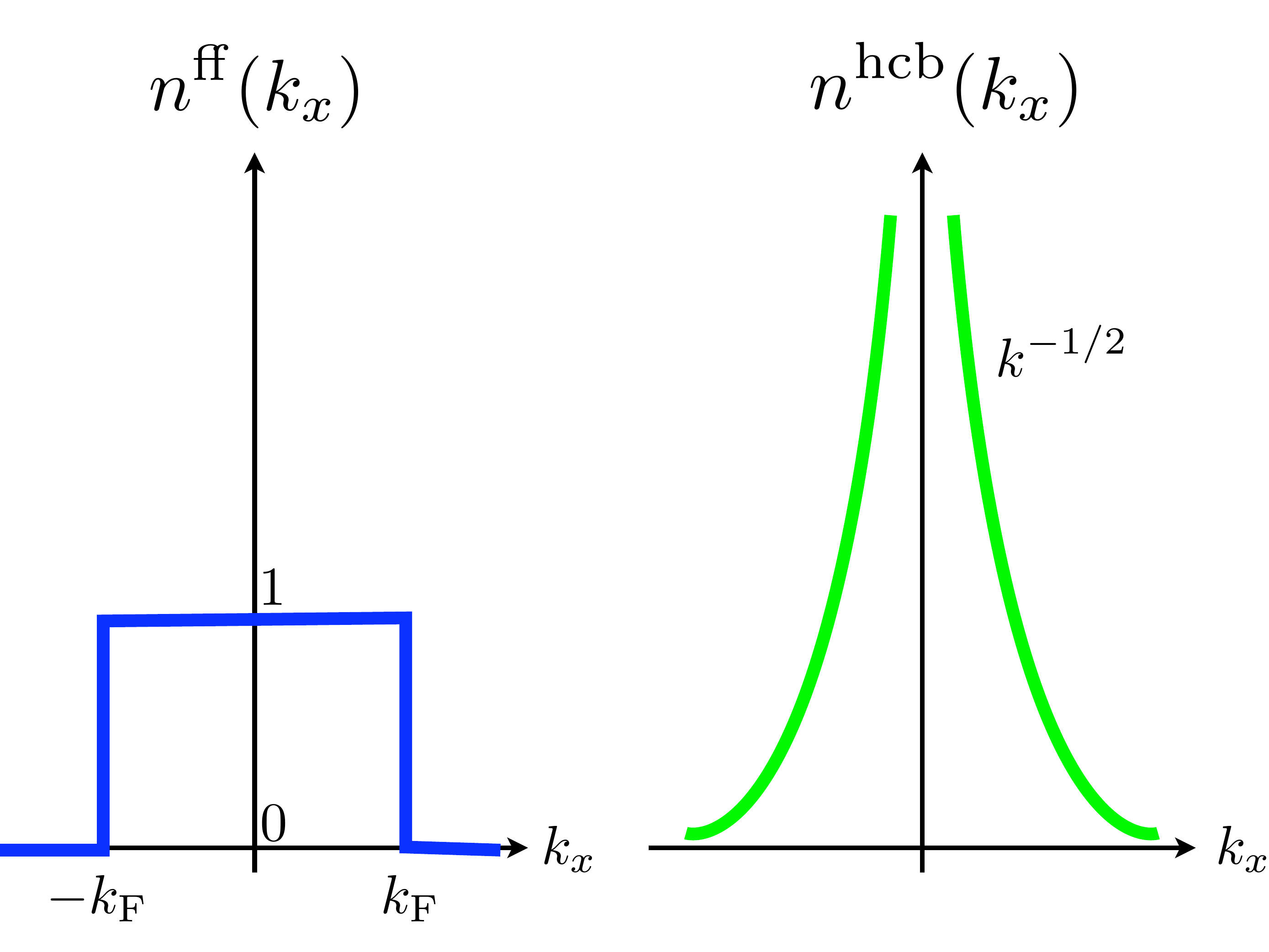}
\caption{\label{fig:nk}(Color online) Schematic representation of the momentum distribution for 1D free-fermions (left) and hard-core bosons (right).}
\end{figure}

Although fermions and hard-core bosons look alike in real space, in terms of their local density, one expect their momentum distribution to be rather different from one another. Indeed the ground-state (GS) of a Fermi gas is the Fermi sea, in which all eigenstates -- momentum states in a free gas -- are filled up to the Fermi energy, whereas interacting bosons exhibit a peak in their the momentum distribution, around $k=0$. Indeed for hard-core bosons at zero-temperature, $\langle{\psi^\dagger(x)\psi(x')}\rangle \sim |x-x'|^{-1/2}$ when $|x-x'|\rightarrow \infty$,\cite{Haldane81} and $n(k) = \int dx dx' e^{ik(x-x')}\langle{\psi^\dagger(x)\psi(x')}\rangle \sim k^{-1/2}$ around $k=0$, as schematized in Fig.~\ref{fig:nk}. This algebraic divergence is a result of the enhancement of quantum fluctuations in reduced dimensions, that forbid the existence of a true condensate.  

\subsection{Free-fermions on a 2-leg ladder: exact solution}

\begin{figure}
\includegraphics[width=\columnwidth,clip]{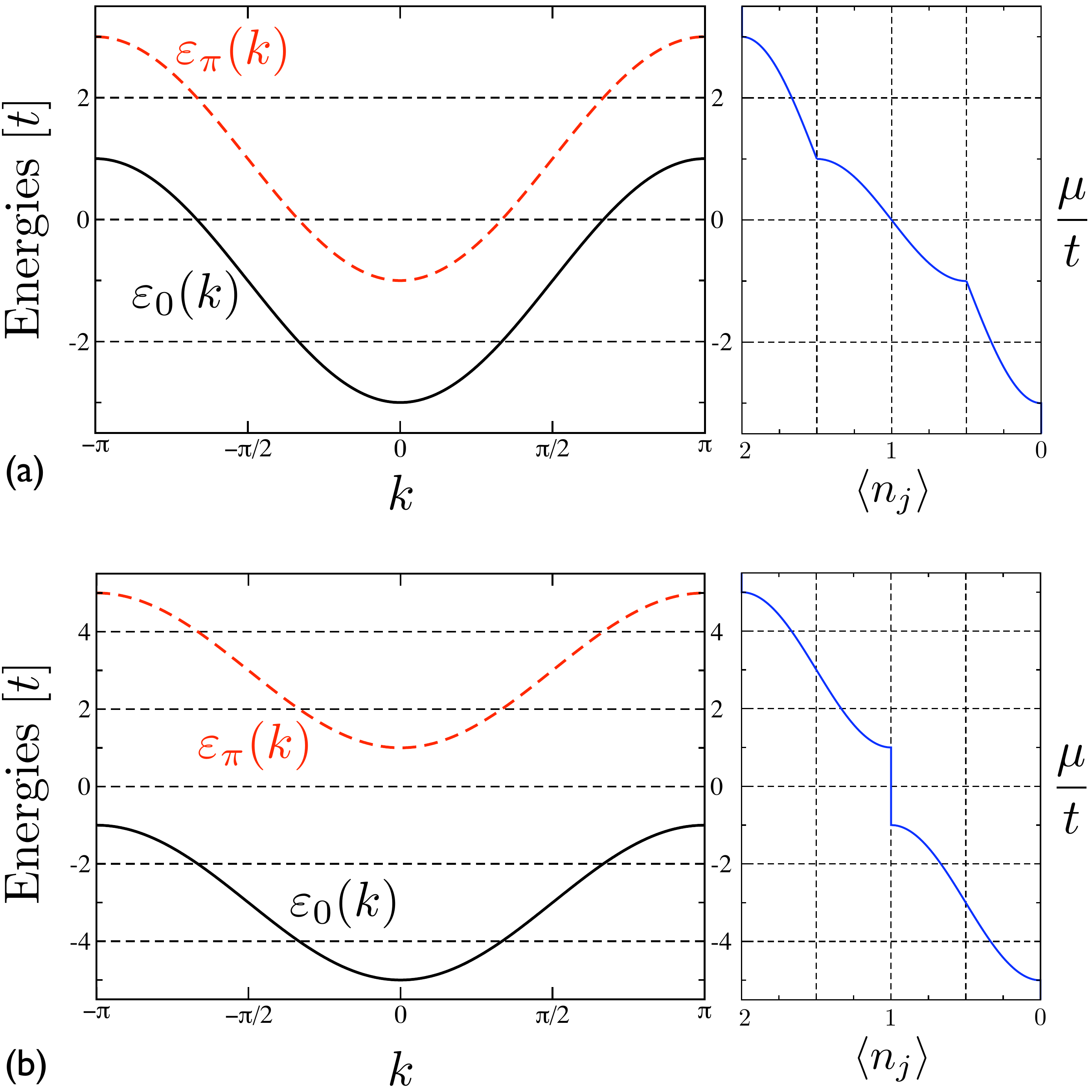}
\caption{\label{fig:band_filling}(Color online) Left: single particle dispersion bands $\varepsilon_0(k)$ (black curves) and $\varepsilon_\pi(k)$ (red dashed) plotted (in units of $t$) for the two-leg ladder model Eq.~(\ref{eq:model}) with $t_{\perp}/t=1$ (a) and $t_{\perp}/t=3$ (b). Right: Corresponding particle filling (per rung) $\langle n_j\rangle =2\rho$ plotted against the chemical potential $\mu/t$.}
\end{figure}

We now turn to the two-leg ladder model Eq.~\eqref{eq:model} in the case of fermionic particles.
Single-particle eigenstates of this Hamiltonian defined on a ladder with $2L$ sites are plane waves of the form:
\begin{eqnarray}
|{k,k_y}\rangle &=& \frac{1}{\sqrt{2L}}\sum_{j=1}^{L} e^{-ik j} \left(|j,1\rangle + e^{ik_y}|j,2\rangle\right)\;,
\end{eqnarray}
where $k$ is a multiple of $2\pi/L$, $k_y=0,\pi$, and $|j,\ell\rangle$ is a state with a single particle at site $(j,\ell)$.
The associated eigenvalues read:
\begin{eqnarray}
\varepsilon_{k_y}(k) &=& -2t\cos(k) -t_\perp\cos(k_y)\,, 
\label{eq:0}
\end{eqnarray}
which corresponds to the band dispersions depicted on Fig.~\ref{fig:band_filling}.
For fermions at zero-temperature, the average filling $\rho$ vs
chemical potential $\mu$ is simply the sum of the fillings of each
band. Two examples for $t_\perp = t$ and $t_\perp = 3t$ are given in
Fig.~\ref{fig:band_filling}. They correspond to two situations. First, if
$t_\perp<2t$, the system has either four Fermi points (when
$-2t+t_\perp < \mu < 2t - t_\perp$) leading to a two-mode LL, or two
Fermi points, leading to a single-mode LL equivalent to a chain. Cusps in the $\rho(\mu)$
curve of Fig.~\ref{fig:band_filling} signal the change in the number
of Fermi points.

Secondly, when $t_\perp>2t$, the two bands are separated by a gap 
\be
\Delta= 2t_\perp -4t\;,
\label{eq:gapff}
\ee 
and, at half-filling $\rho=1/2$, the system is a band insulator. The
corresponding GS has one particle on each rung in the symmetric state
$(|j,1\rangle + |j,2\rangle)/\sqrt{2}$ with a total energy $E_{\rm GS}
= -Lt_\perp$. Away from the half-filling, the ladder is in a single-mode LL.

The situation for hard-core bosons cannot be simply understood in
terms of filling single-particle states. A for a single chain, a
Jordan-Wigner transformation can be performed by choosing a particular
path for the string along the lattice~\cite{Azzouz1994, Dai1998,
  Hori2004}. However, the fermionic model thus obtained cannot be
solved exactly as it contains correlated-like hopping
terms. Mean-field approximations yield reasonable but non-quantitative
results, underlying the fact that the bosonic version of the model is
non-trivial. In the following, we derive quantitatively exact results
working in the bosonic language.

\section{Analytical results for hard-core bosons}
\label{sec:strong_rung}
\subsection{Strong coupling limit}
\subsubsection{Gap at half-filling}
\label{sec:strong_rung_gap}
\begin{figure}
\includegraphics[width=\columnwidth]{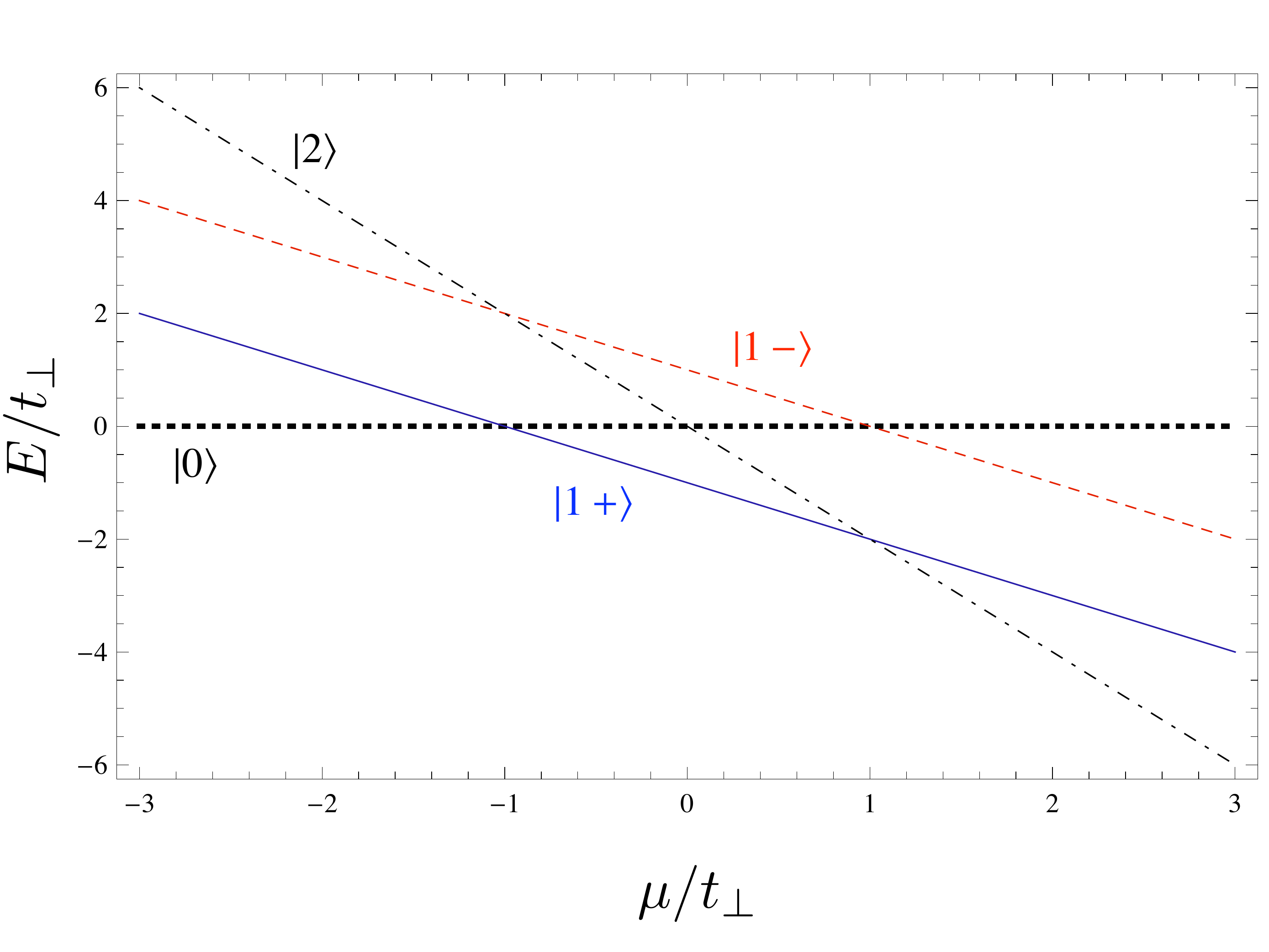}
\caption{(Color online) Energies of the four available states on a given rung, in the limit $t/t_\perp \rightarrow 0$.}
\label{Fig:energies}
\end{figure}

In this section, we compute the gap at half-filling in the large-$t_\perp$ limit, in which $t_\perp$ corresponds to the energy cost to add one extra hard-core boson on top of a half-filled ladder. Details of the calculation can be found in appendix~\ref{app:1}. Setting $t=0$, rungs decouple and four states are available on each of them: an empty state $|0\rangle_j$ with energy $E_0 = 0$, two 1-particle states $|1\pm\rangle_j = \frac{1}{\sqrt{2}}\left(|j,1\rangle \pm |j,2\rangle\right)$ with energies $E_{1\pm} = \mp t_\perp - \mu$, and a 2-particle state $|2\rangle_j$ with energy $E_2 = -2\mu$. These energies are plotted against the chemical potential $\mu$ in Fig. \ref{Fig:energies}. The average filling $\rho$ of the lattice is $0$ as long as $\mu<-t_\perp$, 1 for  $-t_\perp < \mu < t_\perp$ and 2 as soon as $\mu> t_\perp$. 

We start from a half-filled ladder of hard-core bosons -- $\mu=0$ and $N=L$ bosons -- and treat $\mathcal{H}_1 = -t\sum_{j,\ell}[b^{\dagger}_{\ell,j}b_{\ell,j+1}^{\dagga}+{\rm h.c.}]$ as a perturbation of $\mathcal{H}_0 = -t_{\perp}\sum_{j}[b^{\dagger}_{1,j}b_{2,j}^{\dagga}+{\rm h.c.}]$. The GS of $\mathcal{H}_0$ is constructed by putting one particle on each rung in the symmetric state $|1+\rangle_j = \frac{1}{\sqrt{2}}\left(|j,1\rangle +|j,2\rangle\right)$. We have $|L^{(0)}\rangle = |1+,1+,\ldots,1+\rangle$.  $\mathcal{H}_1$ creates $2L$ {\it particle-hole excitations} on neighboring rungs and induces a second-order correction to the GS energy: 
\be
E_{L} = -t_\perp L -t^2L/t_\perp.
\ee
The first excited states of $\mathcal{H}_0$ with $N+1$ particles are $L$ times degenerate and best written in the momentum representation:
\begin{equation} 
|(L+1)^{(0)}_k\rangle= \frac{1}{\sqrt{L}}\sum_{j} e^{ikj} |1+,\ldots,1+,\underset{(j)}{2},1+,\ldots,1+\rangle. 
\end{equation}
$\mathcal{H}_1$ lifts the degeneracy : indeed, it moves the extra particle along the ladder with both nearest (first order perturbation) and next nearest (second order perturbation) hopping and creates $2(L-2)$ particle-hole excitations on nearest-neighboring rungs. Therefore the corrected energies $E_{L+1}(k)$ of the $L+1$-particle states are 
\bea
E_{L+1}(k) &=& -t_\perp(L-1) -2t\cos(k)\nonumber \\
 &-&(t^2/t_\perp)\cos(2k) -(L-2)t^2/t_\perp.
\eea
The situation and calculation are similar when doping the half-filled ladder with holes. The size of the plateau at half-filling is then
\be
\Delta_s = 2[E_{L+1}(0)-E_{L}] = 2t_\perp - 4t + 2t^2/t_\perp.
\label{eq:gapsc}
\ee 
The gap is larger in the case of hard-core bosons than it is in the case of spinless fermions. To conclude this section, it is instructive to look at the perturbative approach in the fermionic case, for which we know that the gap is exactly $\Delta_s = 2t_\perp - 4t$, without a second order correction. This is due to the fact that the half-filled GS $|L\rangle = c^\dagger_{+,0}\ldots c^\dagger_{+,L-1}|0\rangle$ and the $L+1$-particle states $|(L+1)_k\rangle = \frac{1}{\sqrt{L}}\sum_j e^{ikja}c^\dagger_{-,j}|L\rangle$ are exact eigenstates of the full Hamiltonian. $H_1$ does not create particle-holes excitations, since $c^\dagger_{+,j}c^\dagger_{+,j} = 0$. On the contrary, in the bosonic case, $b^\dagger_{+,j}b^\dagger_{+,j} = \frac{1}{2}(b^\dagger_{1,j}b^\dagger_{1,j}+b^\dagger_{2,j}b^\dagger_{2,j}) + \frac{1}{2}(b^\dagger_{1,j}b^\dagger_{2,j}+b^\dagger_{2,j}b^\dagger_{1,j})$. The diagonal terms vanishes because of the hard-core constraint, while the off-diagonal terms are equal because of the commutation of the bosonic operators (they vanish for fermions!). This difference, $c^\dagger_{+,j}c^\dagger_{+,j} = 0$ versus $b^\dagger_{+,j}b^\dagger_{+,j} = b^\dagger_{1,j}b^\dagger_{2,j}$, illustrates how the simplest quasi-1D system is indeed sensitive to quantum statistics.

%
\subsubsection{Effective model for incommensurate fillings}
\label{sec:EffMod}
We can analyze the strong rung coupling limit in the same spirit as for Heisenberg spin ladders in a magnetic field~\cite{Mila98,Totsuka98}. According to Fig.~\ref{Fig:energies}, at a critical value of the chemical potential $\mu_c =t_\perp$, levels $|1+\rangle_j$ and $|2\rangle_j$ become degenerate, with an energy $E_0=-2t_\perp$. Doping the plateau with particles can be studied through degenerate perturbation theory on the Hamiltonian
\begin{equation}
\mathcal{H}_1 = -t\sum_{j,\ell}\big[b^{\dagger}_{\ell,j}b_{\ell,j+1}^{\dagga}+{\rm h.c.}\big] -(\mu - \mu_c) \sum_{j,l} b^\dagger_{j,\ell}b^\dagga_{j,\ell},
\end{equation}
with respect to the GS Hamiltonian
\begin{equation}
\mathcal{H}_0 = -t_{\perp}\sum_{j}\big[b^{\dagger}_{1,j}b_{2,j}^{\dagga}+{\rm h.c.}\big]- \mu_c \sum_{j,l} b^\dagger_{j,\ell}b^\dagga_{j,\ell}.
\end{equation} 
Let us call $P_0$ the projector on the GS subspace, in which $|1+\rangle_j$ and $|2\rangle_j$ are the only states allowed on a given rung $j$. We call $Q_0$ the projector on the complementary subspace. The effective Hamiltonian to second order in perturbation theory is\cite{Messiah}
\begin{equation}
{\cal{H}}_{\rm eff} = P_0 {\cal{H}}_1 P_0 + P_0 {\cal{H}}_1 Q_0 [E-H_0]^{-1} Q_0 {\cal{H}}_1 P_0,
\end{equation}
with $E$ the eigenvalue of the degenerate subspace under consideration. Here, given the degenerate GS and the form of ${\cal{H}}_1$, a {\it virtual} excited state is a state with an empty rung. Then, in everything that follows, $Q_0[E-H_0]^{-1}Q_0 = Q_0[-2t_\perp L + 2t_\perp(L-1)]^{-1}Q_0 = -Q_0/(2t_\perp)$. Again, a more detailed calculation is given in appendix~\ref{app:1}. We find the following effective Hamiltonian expressed in terms of spinless fermions:
\begin{eqnarray}
{\cal{H}}_{\rm eff} &=& -t\sum_j\left[ c^\dagger_j c^\dagga_{j+1} + {\rm h.c} \right] - (\mu-\mu_c)\sum_j (n_j+1)\nonumber\\
&&-\frac{t^2}{2t_\perp}\sum_j \left[ c^\dagger_{j-1} (1-n_{j}) c^\dagga_{j+1} + {\rm h.c} \right]\nonumber\\
&&-\frac{t^2}{t_\perp}\sum_j (1-n_j)(1-n_{j+1}) 
\label{eq:Hefffermions}
\end{eqnarray}
In this language, a spinless fermion stands for a doubly occupied rung, while an empty site indicates a rung with one particle in the symmetric state. Note that one recovers the expression of the gap to second order. One particle is added when the effective chemical potential equates the bottom of the energy band: $\mu-\mu_c -2t^2/t_\perp = -2t -t^2/t_\perp$. On the other side of the plateau the exact same Hamiltonian is obtained, with spinless fermions now standing for empty rungs, and $\mu_c = -t_\perp$.

As already noticed in the context of 2D supersolid phases of quantum antiferromagnets in a field~\cite{Schmidt08,Picon08,Albuquerque10} or Heisenberg ladders in a field~\cite{Bouillot11}, an important ingredient which emerges as a second order process in the above effective Hamiltonian Eq.~\eqref{eq:Hefffermions} is the correlated (or assisted) hopping between second neighbors. For Bose-Hubbard chains in the limit of large but finite on-site repulsion, an effective model of spinless fermions similar to Eq.~\eqref{eq:Hefffermions} was obtained~\cite{Cazalilla03}. We discuss further the relationship with Bose-Hubbard chains and spin ladders in an external field below in sections~\ref{sec:DBG} and \ref{sec:exp}.

%
%
%
\subsection{Weakly coupled chains: bosonization}
\label{sec:bos}
\begin{figure}
\includegraphics[width=\columnwidth,clip]{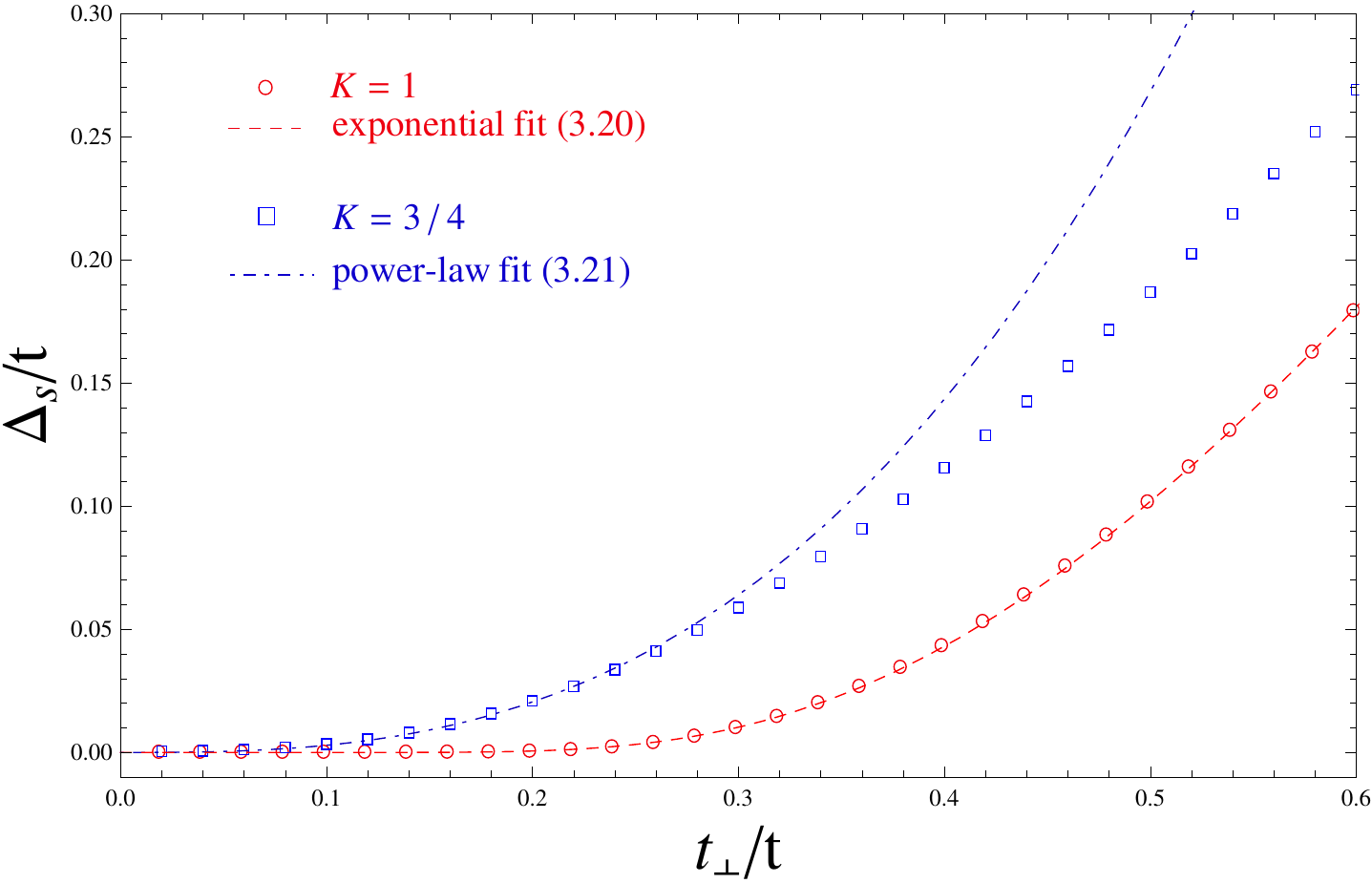}
\caption{\label{fig:gap_RG}(Color online) Charge gap obtained from the numerical integration of the two-step RG flow. The scaling of the gap depends strongly on the initial value of $K$. We find for the exponent of the power-law, eq. \eqref{eq:RGpowerlaw}, $b=2.81$ in good agreement with the DMRG result of equation \eqref{eq:gapXXZ}.}\end{figure}

In this section, we develop a low energy approach in order to analyze the behavior of two weakly coupled chains of interacting bosons at half filling.
Our starting point is the following Hamiltonian ${\cal H}={\cal H}_0+\cal H_\perp$ with
\bea
{\cal H}_0 &=& - t \sum_{j,\ell} \big[ b^\dagger_{\ell,j+1}b_{\ell,j}^{\dagga} +{\rm h.c.}\big]
+\frac{U}{2}\sum_{j,\ell} n_{\ell,j}(n_{\ell,j}-1)\,,\nonumber\\
\label{eq:hb0}
{\cal H_\perp}&=&- t_\perp\sum_{j} \big[ b^\dagger_{1,j}b^\dagga_{2,j} +{\rm h.c.}\big]\,.
\eea
${\cal H}_0$ describes two uncoupled chains of bosons interacting via the onsite repulsion $U$ such that Eq.~\eqref{eq:model} is recovered for $U=\infty$.
$\cal H_\perp$ couples the two chains. The low-energy excitations of a single chain of interacting bosons are collective excitations corresponding to the sound modes in the quasi-condensate. A suitable approach to describe them is bosonization~\cite{Giamarchi-book}. Within this representation,
we can write a bosonic creation operator $\psi_b^\dag(x)$ ($\psi_b^\dag(x)$ being the continuum limit  version of the lattice boson creation operator $b^\dag$) as
\begin{equation}
\label{eq:dictionary}
\psi_b^\dag(x)=\left[\rho -\frac{1}{\pi}\nabla\Phi \right]^{1/2}\sum_p e^{i2p(\pi\rho x-\Phi(x))}e^{-i \theta(x)},
\end{equation}
where $p$ is an integer and $\rho$ is the boson density. $\Phi$ and $\theta$ are two dual bosonic fields
obeying the following commutation relation
\begin{equation}
\Big[\Phi(x),\frac{1}{\pi}\nabla\theta(x')\Big]=i\delta(x-x').
\end{equation}
In this language, $-\frac{1}{\pi}\nabla\Phi$ accounts for the long wavelength density fluctuations, while $\theta$ can be regarded as the superfluid phase. Using Eq. (\ref{eq:dictionary}) and taking the continuum limit, we can rewrite ${\cal H}_0$ in Eq. (\ref{eq:hb0})
as \cite{Giamarchi-book}
\begin{equation}
H_0=\frac{v}{2\pi}\sum_{\ell=1,2}\int dx\, \left[K(\nabla\theta_\ell)^2+\frac{1}{K}(\nabla\Phi_\ell)^2\right],
\end{equation}
where we have introduced the pair of bosonic fields $\Phi_\ell$ and $\theta_\ell$ living on chain $\ell$. 
The low-energy physics of each chain is characterized by the same two parameters, $v$ the sound velocity and
$K$ the Luttinger parameter which encodes interactions in the system.
For interacting bosons  with on-site repulsion, $K\geq 1$.  With nearest-neighbor interactions it is possible to reach $K \leq 1$, thus making a connection with XXZ spin chains. Using Eq. (\ref{eq:dictionary}) we can rewrite $\cal H_\perp$ as
\begin{eqnarray}
\label{eq:hperp}
{\cal H}_\perp&\approx& -\frac{t_\perp}{\pi\alpha}\int dx
\cos(\theta_2-\theta_1)\times\\
&&\Bigl[1 
+4\cos(2\pi\rho x-\Phi_2-\Phi_1)\cos(\Phi_2-\Phi_1)\nonumber\\
\nonumber
&&+2\cos(4\pi\rho x-2\Phi_2-2\Phi_1)+2\cos(2\Phi_2-2\Phi_1)\,\Bigr].
\end{eqnarray}
In order to write this equation, we restricted the summation in Eq. (\ref{eq:dictionary}) to the values $p=0,1,-1$ which provide the leading terms in the perturbation. Half-filling corresponds to $\rho=1/(2a)$ with $a$ the lattice spacing.
Introducing symmetric and antisymmetric fields defined as
$\Phi_{s}=(\Phi_1+\Phi_2)/\sqrt{2}$,$\Phi_{a}=(\Phi_1-\Phi_2)/\sqrt{2}$, $\theta_{s}=(\theta_1+\theta_2)/\sqrt{2}$, $\theta_{a}=(\theta_1-\theta_2)/\sqrt{2}$  
and retaining only the non-oscillatory terms, we can rewrite ${\cal H=H}_0+\cal H_\perp$ as
\begin{eqnarray}\label{eq:hbos}
{\cal H}&=& \sum_{j=s,a} \frac{v_j}{2\pi}\int dx \,\Big[K_j(\nabla\theta_j)^2
+\frac{1}{K_j}(\nabla\Phi_j)^2\Big] \nonumber\\
&&-\frac{t_\perp}{\pi\alpha}\int dx\,\left[
\cos(\sqrt{2}\theta_a)+
2\cos(\sqrt{2}\theta_a)\cos(\sqrt{8}\Phi_s)\right.\nonumber\\
&&\qquad\qquad\quad\left.+2\cos(\sqrt{2}\theta_a)\cos(\sqrt{8}\Phi_a) \right],
\end{eqnarray}
where $K_a=K_s=K$ and $v_a=v_s=v$. We see that the coupling between the two chains is quite complex in the bosonized representation. To study this kind of Hamiltonian a strategy consists in deriving RG equations for the various coupling constants. To do so, we first write the Euclidian action $S$ associated to Eq. (\ref{eq:hbos}) in the following general form:
\begin{eqnarray}\label{eq:sbos}
S&=& \sum_{j=s,a} \frac{1}{2\pi K_j}\int dx d\tau\,\Big[\frac{1}{v_j}(\partial_\tau\Phi_j)^2+v_j(\partial_x\theta_j)^2\Big] \nonumber\\
&&-\int dx d (v\tau) \,\Bigl[\frac{y_\perp}{\pi a^2}\cos(\sqrt{2}\theta_a)\nonumber\\
&&\qquad\qquad+\frac{2 y_a}{\pi a^2}\cos(\sqrt{2}\theta_a)\cos(\sqrt{8}\Phi_a)\nonumber\\
&&\qquad\qquad+\frac{2 y_s}{\pi a^2}\cos(\sqrt{2}\theta_a)\cos(\sqrt{8}\Phi_s) \nonumber\\
&&\qquad\qquad+\frac{ g_s}{\pi a^2}\cos(\sqrt{8}\Phi_s)\Bigr]\;,
\end{eqnarray}
where we have introduced four dimensionless coupling constants 
$y_\perp$, $y_s$, $y_a$ and finally $g_s$ which is generated under the RG flow as we will see. These couplings constants have bare values which are directly extracted from the Hamiltonian in  Eq. (\ref{eq:hbos}) and read 
$y_\perp(a)=a t_\perp/ v$, $y_s(a)=y_a(a)=a t_\perp v$, $g_s(a)=0$. 
We write RG equations 
for the six coupling constants using standard techniques~\cite{Giamarchi-book}:

\begin{eqnarray}\label{eq:rggen}
\frac{dy_\perp}{dl}&=&\Big(2-\frac{1}{2K_a}\Big)y_\perp(l),\nonumber\\
\frac{dy_a}{dl}&=&\Big(2-\frac{1}{2K_a}-2K_a\Big)y_a(l),\nonumber\\
\frac{dy_s}{dl}&=&\Big(2-\frac{1}{2K_a}-2K_s\Big)y_s(l),\\
\frac{dg_s}{dl}&=&(2-2K_s)g_s(l)+\frac{1}{\pi}y_\perp(l) y_s(l),\nonumber\\
\frac{dK_a}{dl}&=&\frac{y_\perp^2(l)}{4}+\frac{y_s^2(l)}{4},\nonumber\\
\frac{dK_s}{dl}&=&-4K_s^2(l)y_s^2(l)-2 g_s^2(l)K_s^2(l),\nonumber
\end{eqnarray}
The running short distance cutoff is parametrized as $a(l)=ae^{l}$. Since the bare values of $K_{s,a}$ are larger than one, the coupling $y_\perp$ is strongly relevant compared to the other couplings and $y_\perp$ is therefore driven to strong coupling. 
This implies that the field $\theta_a$ becomes gapped. The superfluid phase fields
 of both chains lock
together as soon as interchain tunneling is switched on. 
Next we introduce the scale $l_\perp$ such that $y_\perp(l_\perp)=1$ which equivalently defines an energy scale
$\Delta_a\sim te^{-l_\perp}$.
For $l<l_\perp$, one can use the full set of RG equations written in Eq.~(\ref{eq:rggen}). For $l> l_\perp$, one can simplify the action~(\ref{eq:sbos}) 
by replacing the field $\cos(\sqrt{2}\theta_a)$ by its average value (a similar procedure is described in Ref.~\onlinecite{Kim99} for coupled spin chains) and thus obtain an 
effective action valid at lower energy scales $l>l_\perp$. The phase field $\theta_a$ being gapped, we focus on the symmetric $(s)$ sector whose effective action $S_s^{\rm eff}$ now takes the simpler form
\bea
S_s^{\rm eff}&=&\int dx d\tau\Big\{ \frac{1}{2\pi \widetilde K_s} \Big[\frac{1}{v_s}(\partial_\tau\Phi_s)^2+v_s(\partial_x\Phi_s)^2\Big]\nonumber\\
&&\qquad\qquad-\frac{ \widetilde g_s}{\pi\alpha^2}\cos(\sqrt{8}\Phi_s)\Big\}\;,
\eea
where we have introduced $\widetilde K_s=K_s(l_\perp)$ and
$\widetilde g_s=g_s(l_\perp)+2\eta y_s(l_\perp)$ with  $\eta=\langle\cos(\sqrt{2}\theta_a)\rangle=O(1)$.
The action $S_s^{\rm eff}$ is of the sine-Gordon type. The RG equations associated to $\widetilde K_s$ and
$\widetilde g_s$ are obtained by further integrating high-energy degrees of freedom between $l_\perp$ and $l$, leading
to
\begin{eqnarray}\label{eq:rg}
\frac{d\widetilde g_s}{dl}&=&\Big(2-2\widetilde K_s(l)\Big)\widetilde g_s(l),\\
\frac{d\widetilde K_s}{dl}&=&-2 \widetilde g_s^2(l)\widetilde K_s^2(l),
\end{eqnarray}
which are the standard Kosterlitz-Thouless flow equations~\cite{Giamarchi-book}.
The coupling $\widetilde g_s$ flows to strong coupling when $2\widetilde K_s(l)$ is smaller than $1$. Since $K_s(l)$ always decreases during the fist stage of the RG transformation, $\widetilde g_s$ is driven to strong coupling and, ultimately, a gap $\Delta_s$ opens up 
in the 
symmetric sector, at a very low energy scale. We stress that the gap opening is non trivial from the RG point of view:  
first, a gap opens in the antisymmetric sector (pining of the field $\theta_a$), which eventually induces the opening of the gap in the symmetric sector (pining of the field $\Phi_s$). The scaling of the gap with $t_\perp$, obtained from the numerical solution of the RG flow, is shown in Fig.~\ref{fig:gap_RG}. It strongly depends on the initial value of $K$. For hard-core bosons -- $K=1$ -- the charge gap grows exponentially slowly with $t_\perp$:
\be
\Delta_s \propto e^{-{\text{a}}t/t_\perp}. 
\ee 
Indeed, during the first step of the flow, $l<l_\perp$, $y_s$ is irrelevant and renormalized downwards to a small value, while $K_s$ decreases to a value smaller but close to 1. At the start of the second step, $y_s$ has become a relevant perturbation but is still very close to marginality. On the contrary we could imagine a situation with $K$ well below 1 -- for instance, adding nearest neighbor repulsive interactions in \eqref{eq:hb0} -- and $y_s$ would much farther away from marginality at the start of the second step, therefore leading to a much faster -- power-law -- opening of the gap (see Fig. \ref{fig:gap_RG}),
\be
\label{eq:RGpowerlaw}
\Delta_s \propto (t_\perp/t)^{b}. 
\ee 
Finally, this whole RG analysis shows that a gap in the symmetric sector opens up as soon as transverse hopping is switched on. This result is drastically different from the fermionic case where a band gap appears only for a finite large value of $t_\perp$.

\subsection{Comparison with spin-$1/2$ ladders}
\label{sec:XXZ}
A two-leg ladder of hard-core bosons is equivalent to a spin-$1/2$ two-leg ladder. Indeed, spins-$1/2$ satisfy mixed commutation relations, when expressed in terms of $S_{\alpha,j}^{\pm} = S_{\alpha,j}^x \pm iS_{\alpha,j}^y$. A hard-core boson on a given site corresponds to a magnetic moment with $m^z=+1/2$ while an empty site corresponds to the opposite with $m^z=-1/2$.  Consider the model of hard-core bosons we have studied and add nearest-neighbour repulsion along and between the chain. It maps to a spin-$1/2$ ladder:
\begin{align}
{\cal H} =& \sum_{j,\ell} J\big[S_{\ell,j}^x S_{\ell,j+1}^x + S_{\ell,j}^y S_{\ell,j+1}^y\big]+ J^z S_{\ell,j}^z S_{\ell,j+1}^z \nonumber \\
     &+ \sum_{j=1}^L J_\perp \big[S_{1,j}^x S_{2,j}^x + S_{1,j}^y S_{2,j}^y\big] + J_\perp^z S_{1,j}^z S_{2,j}^z \label{XXZ_ladder_hamiltonian}\,.
\end{align}
For $J^z = J^z_\perp = 0$ it reduces to the model Eq.~(\ref{eq:model}), with $t=J/2$ and $t_\perp=J_\perp/2$.  The strong coupling analysis\cite{Barnes93,Mila98,Totsuka98} leads to a similar picture: singlets form on each rung and a finite spin gap ($m=0$ magnetization plateau) extends around zero magnetic field. A weak-coupling analysis leads to the same conclusion. However the mechanism for the formation of the plateau is different than in our study of hard-core bosons because of the $J_\perp^z$ exchange coupling. We recall below the bosonized form of the Hamiltonian (\ref{XXZ_ladder_hamiltonian}). In a similar way to the physics described in section \ref{sec:bos}, the dynamics decouple into two modes, symmetric and antisymmetric such that one can write ${\cal H = H}_s + {\cal H}_a$~\cite{Strong92} with
\begin{eqnarray}
{\cal H}_s &=& \frac{v}{2\pi}\int dx\, \left[K_s(\nabla\theta_s)^2+\frac{1}{K_s}(\nabla\Phi_s)^2\right]\nonumber\\
&&+ \frac{2g_s}{(2\pi a)^2}\int dx \cos[\sqrt{8}\Phi_s],\\
{\cal H}_a &=& \frac{v}{2\pi}\int dx\, \left[K_a(\nabla\theta_a)^2+\frac{1}{K_a}(\nabla\Phi_a)^2\right]\nonumber\\
&&+ \frac{2g_a}{(2\pi a)^2}\int dx \cos[\sqrt{8}\Phi_a]\nonumber\\
&& + \frac{2G_a}{(2\pi a)^2}\int dx \cos[\sqrt{8}\theta_a].
\end{eqnarray}
Here, $K_s = K\left(1-\frac{KJ^z_\perp a}{2\pi v}\right),  K_a = K\left(1+\frac{KJ^z_\perp a}{2\pi v}\right), g_s = g_a = a J^z_\perp , G_a = \pi J_\perp a$. $K$ is the original Luttinger parameter of the single spin chain and depends on $J^z$, the anisotropy parameter. For instance $K=1$ for $J^z=0$ (hard-core bosons) and $K=1/2$ for $J^z=J$ (Heisenberg chain). Note that the $z$ exchange along the chains should also give rise to an umklapp term of the form:
\begin{equation}
{\cal H}_u =  \frac{2g_u}{(2\pi a)^2}\int dx \cos[\sqrt{8}\Phi_s] \cos[\sqrt{8}\Phi_a]
\end{equation}
with $g_u \propto J^z$. However as long as $K>1/2$ this term is always less relevant than the other interaction terms and will not open any gap. When $J^z_\perp$ vanishes one should also include the term $\cos[\sqrt{2}\theta_a]\cos[2\sqrt{2}\Phi_s]$ which we studied in the previous section. For a non-zero $J^z_\perp$, it is anyway less relevant and the gap in the symmetric sector opens directly because of the exchange along $z$. Indeed, for all $J^z\leq 1$ and $J^z_\perp \neq 0$, $K_s<1$ and $g_s$ is relevant, therefore ordering the field $\Phi_s$. The scaling of the gap is known from the sine-Gordon model and depends on the initial conditions for $K_s$ and $g_s$. Away from marginality, the charge gap scales as a power-law:
\begin{equation}
\Delta_s \simeq \left( J^z_\perp\right)^{\frac{1}{2-2K}}\,.
\end{equation}
For instance, at the isotropic Heisenberg point $J^z = J$, $K=1/2$ and the gap opens linearly with $J_\perp$.

Since the antisymmetric sector is always gapped, we expect the familiar picture of "rung-singlet" known for Heisenberg ladders~\cite{Dagotto96} to hold here for the generic U(1) symmetric case for which the gapped state is called a {\it{Rung-Mott-Insulator}} (RMI). However, the correlation length of the symmetric mode $\xi\sim 1/\Delta_s$ is much larger than for Heisenberg ladders. Therefore, the usual short-range resonating singlets picture, very useful for spin ladders~\cite{White94,Dagotto96}, has to be taken more carefully here, especially in the limit of small $t_\perp/t$. 

\section{Rung-Mott Insulator: Numerical results}
\label{sec:RMI}
Coming back to the original problem of hard-core bosons, we present in this section quantum Monte Carlo~\cite{Sandvik91,Syljuaasen02,Evertz03} (QMC) and density-matrix renormalization group~\cite{White1992, White1993, Schollwock2005} (DMRG) results at half-filling. The Rung-Mott insulating state (RMI) will be studied in great details using these two techniques.
\subsection{Density plateau at half-filling}
\subsubsection{QMC results}
We use the QMC stochastic series expansion (SSE) algorithm~\cite{Sandvik91}, in its directed loop framework~\cite{Sandvik99,Syljuaasen02}. This method has been used quite intensively for the last decade. For our purpose here, the implementation of the algorithm is rather straightforward since one can, for simplicity, work in the $S^z$ representation of the equivalent spin-$1/2$ XY model in a transverse field. Indeed, using the Matsubara-Matsuda mapping~\cite{Matsubara56}, hard-core bosons operators are replaced by spin-$\frac{1}{2}$ operators:
\bea
{\cal H}&=&
-2t\sum_{j,\ell=1,2}\left(S^{x}_{\ell,j}S_{\ell,j+1}^{x}+S^{y}_{\ell,j}S_{\ell,j+1}^{y}\right)-\mu\sum_{j,\ell=1,2}S^{z}_{\ell,j}\nonumber\\
&&-2t_{\perp}\sum_{j}\left(S^{x}_{1,j}S_{2,j}^{x}+S^{y}_{1,j}S_{2,j}^{y}\right)+{\rm{const.}}
\label{eq:modelXY}
\eea
This equivalence with spin-$1/2$ systems, already evoked in section~\ref{sec:XXZ} to discuss the connection with XXZ spin ladders, will also be used later in section~\ref{sec:disc} when discussing potential links of our results with experiments on magnetic field induced LL phases in spin ladders.

Coming back to the bosonic language, we work in the grand-canonical ensemble where the total particle number (longitudinal magnetization) can fluctuate. We measure the following observables: (i) The density of particles
\be
\rho=\frac{1}{2L}\sum_{j,\ell}\langle n_{j,\ell}\rangle,
\ee
(ii) the compressibility
\be
\kappa=\frac{\beta}{2L}\Big[\Big\langle\Big(\sum_{j,\ell} n_{j,\ell}\Big)^2\Big\rangle -\Big\langle\sum_{j,\ell} n_{j,\ell} \Big\rangle^2\Big],
\label{eq:kappa}
\ee
and (iii) the superfluid stiffness
\be
\Upsilon_{\rm sf}=\left.\frac{1}{2L}\frac{\partial^2 E_0(\varphi)}{\partial \varphi^2}\right\vert_{\varphi=0}.
\ee
In the above definitions, $\beta$ is simply the inverse temperature $k_{\rm B} T$, and $\varphi$ is a small twist angle enforced on all bonds in the direction~\cite{Yang1961,Shastry90} along the legs. Technically, the superfluid stiffness~\cite{Fisher73} is efficiently measured via the fluctuations of the winding number~\cite{Pollock87} during the SSE simulation~\cite{Sandvik97}. 
\begin{figure}
\includegraphics[width=\columnwidth,clip]{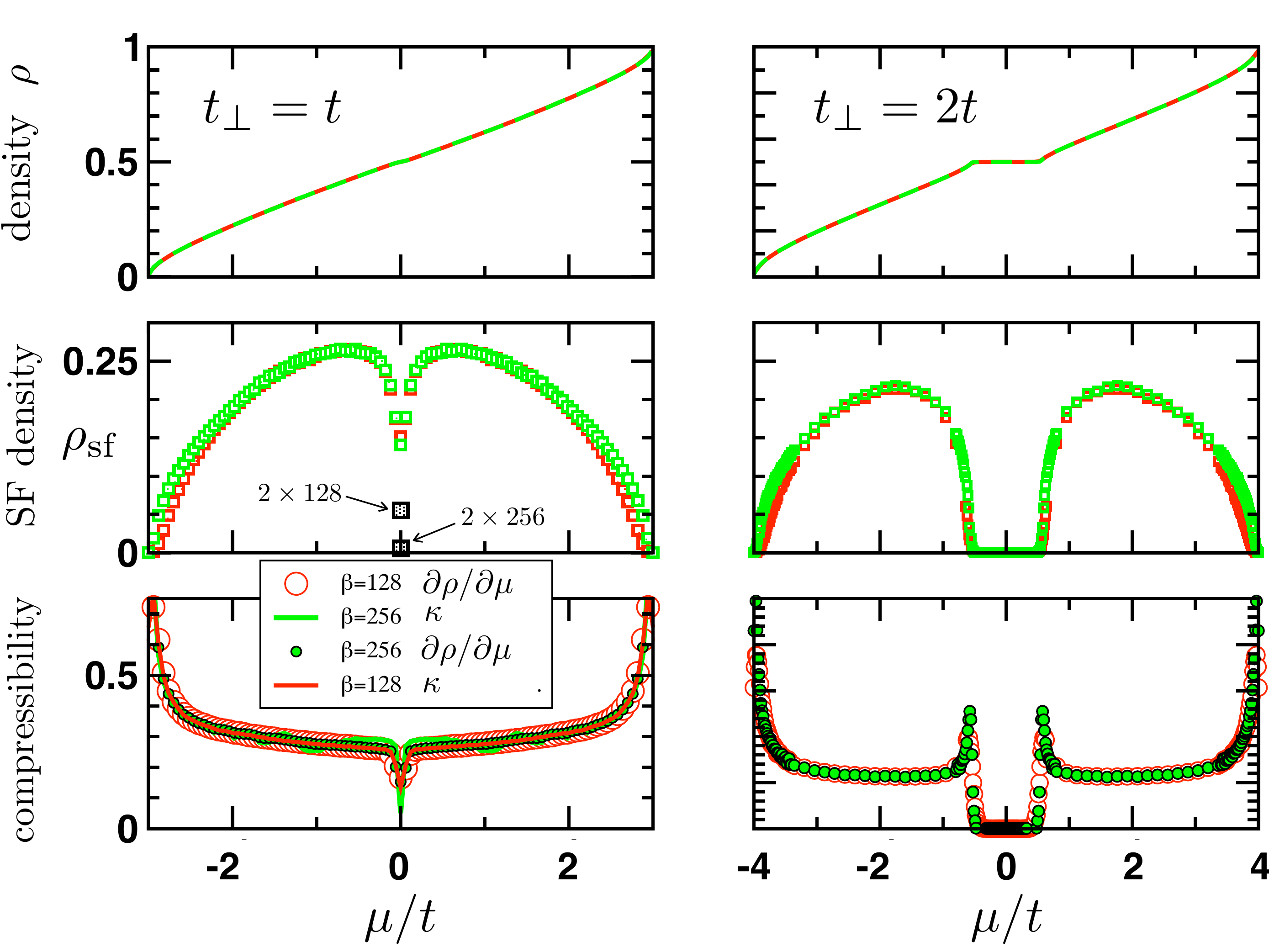}
\caption{\label{fig:rho_and_kappa}
(Color online) QMC results for the particle density $\rho$ (top), the SF density $\rho_{\rm sf}$ (middle), and the compressibility $\kappa$ (bottom) obtained on $2\times 64$ ladders with $t_{\perp}=t=1$ (Left) and $t_{\perp}=2t$ (Right), for two inverse temperature $\beta=128, 256$, as indicated on the plot. On the left panel ($t_\bot=t$), the compressibility data are obtained from either a direct measurement of $\kappa$ [full lines Eq.~(\ref{eq:kappa})], or from a numerical derivative of $\rho(\mu)$ [circles Eq.~(\ref{eq:drhodmu})]. Note for $t_\perp=t$ (left) the superfluid density data are also shown for $L=128$ and $L=256$ at $\beta=4\times L$.}
\end{figure}
In Fig.~\ref{fig:rho_and_kappa}, we show QMC results for the total particle density $\rho$, the superfluid density $\rho_{\rm sf}$ (directly related to $\Upsilon_{\rm sf}$ as discussed below in Sec.~\ref{sec:SF}), and the compressibility $\kappa$. Note that the later can also be extracted from the numerical derivative
\be
\kappa(\mu)=\frac{\partial\rho}{\partial\mu}.
\label{eq:drhodmu}
\ee
QMC data are shown for two representative values of the interchain hopping: $t_\perp=t$ and $t_\perp=2t$. As expected for large $t_\perp$, a plateau in the density at half-filling is clearly visible for $t_\perp=2t$ if $-0.5\lesssim \mu/t\lesssim 0.5$, signaling the incompressible RMI state where the superfluid density and the compressibility both vanish. For $t_\perp=t$, interpretation of finite size numerical data is more difficult since the gap at half-filling turns out to be much smaller. Indeed, the density curve versus $\mu$ does not show any visible plateau (top left of Fig.~\ref{fig:rho_and_kappa}). Nevertheless, the gap tends to show up in the compressibility $\kappa$ which displays a downward feature, signaling the insulating state. The SF density also vanishes, as it becomes visible when $L$ is increased from $64$ up to $256$.
Obviously, when $t_\perp/t$ gets smaller, it will be quite difficult to directly get quantitative information from $\rho$ or $\kappa$, like for instance the value of the gap $\Delta_s$ at half-filling. If we want to do so, one needs to run QMC simulations in the GS, i.e. at temperature well below the energy gap $T\ll \Delta_s$ {\it{and}} in the thermodynamic limit, i.e. for system lengths $L\gg \xi$ where $\xi\sim1/\Delta_s$ is the correlation length associated with the short-range order of the RMI. Of course this would be a difficult task, that we can fortunately circumvent. 

In a QMC framework, the most useful observable to characterize the insulating state, and to get a precise estimate of the correlation length $\xi$ is the superfluid stiffness computed directly at half-filling, which is expected to vanish as $\exp (-L/\xi)$. A direct estimate of the gap with SSE, in principle possible, would require a much more important numerical effort. Zero-temperature DMRG simulations are more competitive to measure tiny gaps as we discuss now before coming back to SSE estimates of the SF stiffness and the correlation length.
\begin{figure}
\centering
\includegraphics[width=\columnwidth,clip]{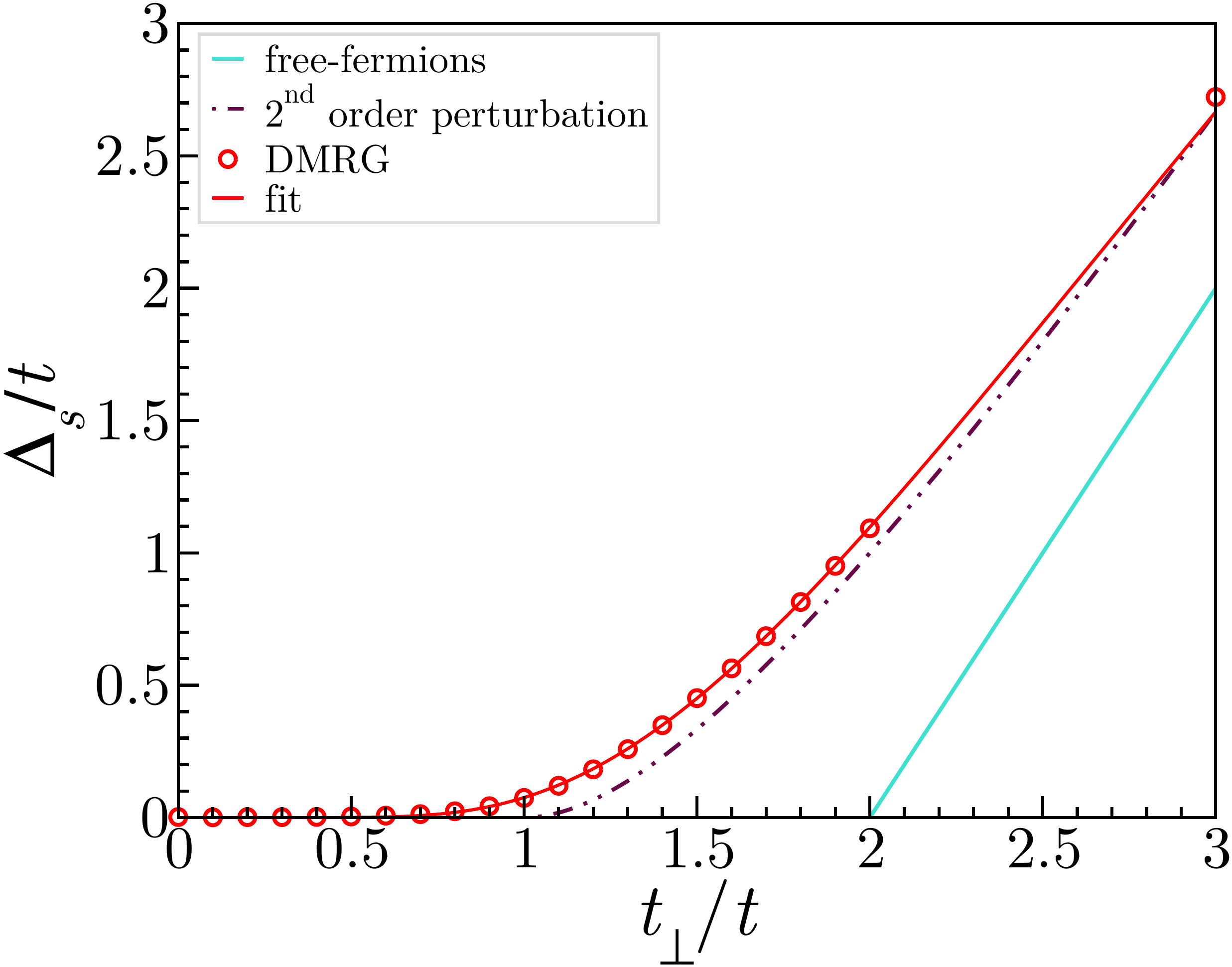}
\caption{\label{fig:gap}(Color online) Charge gap of the two-leg
  ladder model as a function of the interchain coupling obtained from
  DMRG calculations, and compared with perturbation theory Eq.~\eqref{eq:gapff} for free-fermions and Eq.~\eqref{eq:gapsc} for the strong rung coupling expansion. The red curve is a fit to the expression Eq~\eqref{eq:gap} (see the text).}
\end{figure}

%
\subsubsection{Gap from DMRG}
The gap is computed at zero temperature with DMRG using the definition
\begin{equation}
\Delta_s = E_0(N+1)+E_0(N-1)-2E_0(N)\;,
\end{equation}
where $E_0(N)$ is the GS energy with $N$ bosons, so that
$\Delta_s$ is directly the width of the plateau. In the DMRG calculation,
we keep 1000 states per block and extrapolate the results over sizes
ranging from 24 to 160 using the following ansatz from the finite-size
scaling of the gap
\begin{equation}
\Delta_s(L) = \Delta_s + \frac A L e^{-L/\xi}\;.
\end{equation}
The result is reported on Fig.~\ref{fig:gap} in which the
strong-coupling of Sec.~\ref{sec:strong_rung_gap} and free-fermions
predictions are also given for comparison. According to the
bosonization prediction of Sec.~\ref{sec:bos}, we fit the opening for
$t_\perp \leq 2t$ using the following law
\be
\Delta_s=\Delta_s^{0}\exp\left(-\text{a}\frac{t}{{t_\perp^{\vphantom{c}}-t_{\perp}^{c}}}\right)\;,
\label{eq:gap}
\ee
with $\Delta_s^0$, a and $t_{\perp}^c$ as fitting parameters. The
obtained values are respectively $\Delta_s^0 \simeq 15.624t$, $\text{a}\simeq 5.294$
and $t_{\perp}^c \simeq 0.0075t$. The latter critical value is
perfectly compatible with $t_{\perp}^{c}=0$ within our numerical
precision. The opening of the gap is particularly slow (non-analytic
in $t_\perp$), but reaches a sizable magnitude $0.074t \lesssim \Delta_s
\lesssim t$ for $t\leq t_\perp\leq 2t$, thus providing a first clear
quantitative difference with free fermions.

\begin{figure}
\centering
\includegraphics[width=\columnwidth,clip]{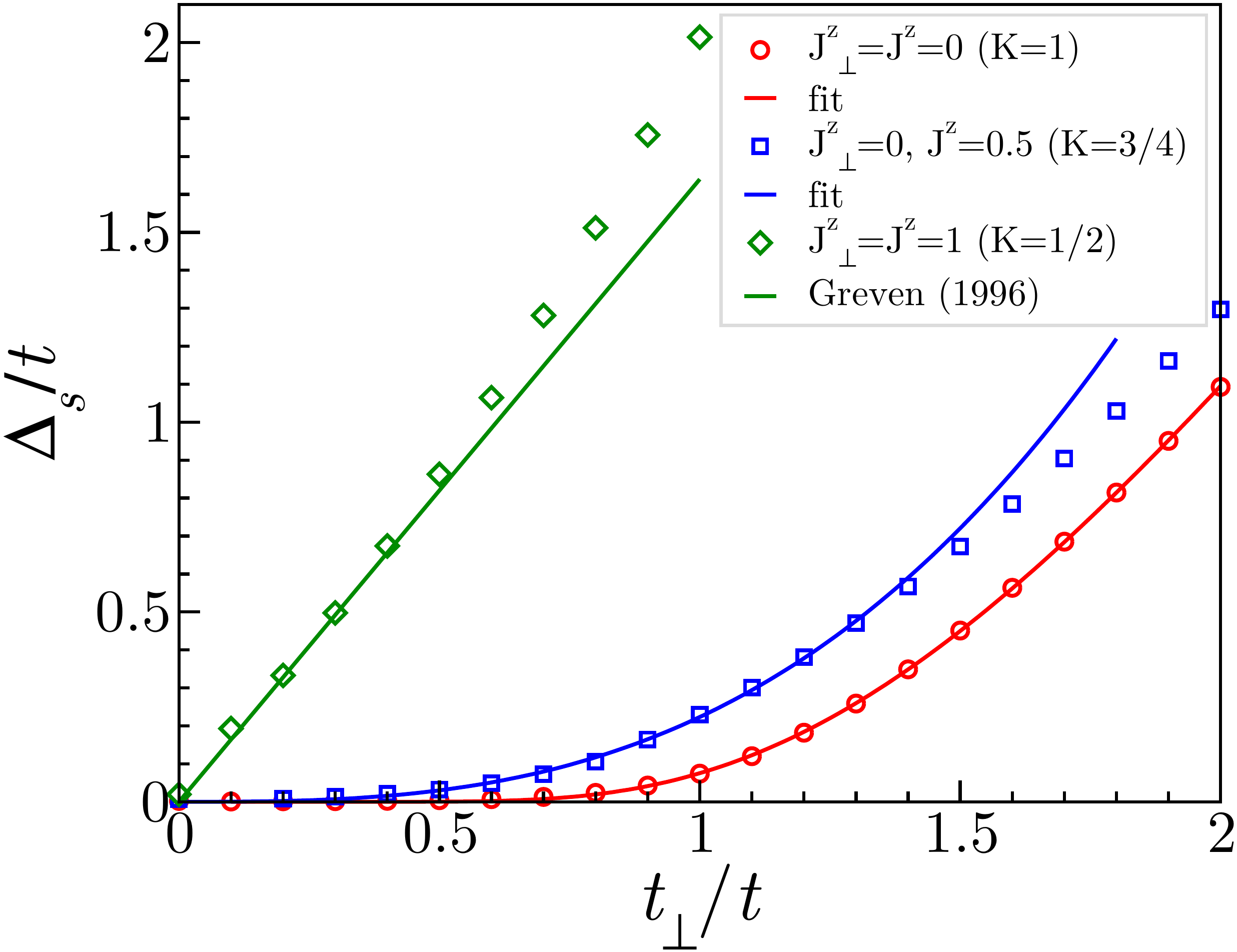}
\caption{\label{fig:gap_comp} (Color online) Comparison of the
  opening of the gap in the XXZ ladder for three different
  situations. 
  DMRG results (symbols) are compared to Eq.~\eqref{eq:gap} for hard-core bosons, to Eq.~\eqref{eq:gapXXZ} for coupled XXZ chains, and to the linear behavior from Greven {\it{et al}}.~\cite{Greven1996} in the SU(2) case.}
\end{figure}

Since it is well known that SU(2) spin-1/2 ladders have a gap at the
isotropic point~\cite{Barnes93} $J=J_\perp$, and that this
gap opens linearly~\cite{Greven1996} with $J_\perp$ in the limit of
weakly coupled chains, we investigate how interactions modify the
opening of the gap in order to have an intermediate situation. We
introduce intra-chain interactions with nearest-neighbor density
interactions corresponding to $J^z=0.5$ in the XXZ
language model. From the Bethe-ansatz solution, the Luttinger
parameter of the chains at zero interchain coupling is thus $K=3/4$.
We also recompute the SU(2) (with
$J^z=J^z_{\perp}=1$ and for which $K=1/2$) gap to
check the numerical accuracy and gather results in
Fig.~\ref{fig:gap_comp}. As expected from the RG calculation of
Sec.~\ref{sec:bos}, the intermediate opening is best fitted by a
power-law with an exponent of the order of three:
\be
\Delta_s \simeq  0.2234\;t \left(\frac{t_\perp}{t}\right)^{2.88} \;,
\label{eq:gapXXZ}
\ee
close to the exponent obtained from the RG calculation, and non-trivially related to the value of $K$ (see Fig.~\ref{fig:gap_RG}).
The gap in the isotropic limit reproduces well the earlier findings of
Ref.~\onlinecite{Greven1996}. The law governing the opening of the gap
is thus very sensitive to interactions, with their expected tendency
to boost the charge gap.

\subsection{Superfluid density}
\label{sec:SF}
\subsubsection{QMC results}

\begin{figure}
\includegraphics[width=\columnwidth,clip]{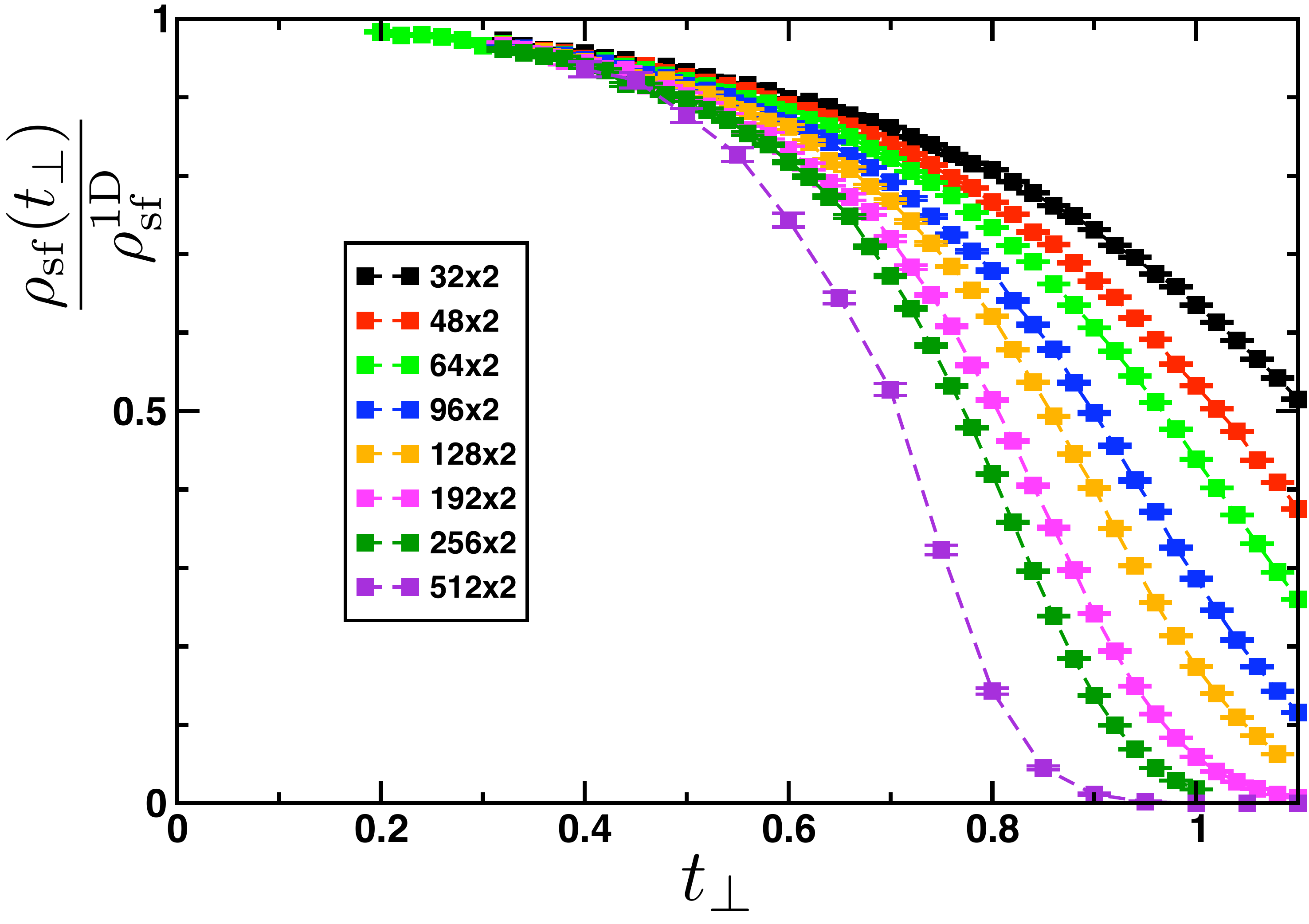}
\caption{\label{fig:rhos_vs_tperp}
(Color online) SF density $\rho_{\rm sf}$ at half-filling (normalized by the decoupled chains value $\rho_{\rm sf}^{\rm 1D}=1/\pi$), plotted versus $t_{\perp}$. QMC results obtained for ladders of size $2\times L$ in the GS at inverse temperatures $\beta=4L$ with $L=32,\cdots,512$.}
\end{figure}
As introduced by Fisher, Barber and Jasnow in Ref.~\onlinecite{Fisher73}, the superfluid stiffness (or helicity modulus) $\Upsilon_{\rm sf}$ is defined by imposing twisted boundary conditions in one direction, such that the hard-core boson Hamiltonian Eq.~(\ref{eq:model}) reads now
\bea
{\cal{H}}(\varphi)&=&
-t\sum_{j,\ell=1,2}\Big[{\rm{e}}^{i\varphi}b^{\dagger}_{\ell,j}b_{\ell,j+1}^{\dagga}+{\rm h.c.}\Big]\\
&&-t_{\perp}\sum_{j}\Big[b^{\dagger}_{1,j}b_{2,j}^{\dagga}+{\rm h.c.}\Big]-\mu\sum_{j,\ell=1,2}n_{\ell,j}.\nonumber
\label{eq:modeltwisted}
\eea
Such a twist angle $\varphi$ mimics the effect of a phase gradient imposed in the bosonic wave function. In a superfluid state, it would lead to a superflow with a gain in the kinetic energy density~\cite{Fisher73}
$\delta E_0(\varphi)=\frac{1}{2}\Upsilon_{\rm sf}\varphi^2=\frac{\hbar^2}{2m^*}\rho_{\rm sf}\varphi^2$, where $m^*$ is the effective mass of the bosons. $m^*$ is determined in the diluted limit where the cosine dispersion $2t\cos k_x$ can be approximatively described as a quadratic free bosons dispersion $\frac{\hbar^2 k_x^2}{2m^*}$, thus leading to $\frac{\hbar^2}{2m^*}=t$, and therefore 
\be
\rho_{\rm sf}=\frac{1}{2L}\frac{1}{2t}\frac{\partial^2 E_0(\varphi)}{\partial \varphi^2}\Big|_{\varphi=0}.
\ee
Based on the winding number fluctuations~\cite{Pollock87,Sandvik97}, we compute for various system sizes the superfluid density at half-filling $\mu=0$ versus interchain hoppings $t_\perp$. As already discussed, it is crucial to get zero-temperature estimates. This is ensured here by performing SSE simulations at $\beta=4\times L$ for the values of $t_\perp$ considered in the following analysis. In Fig.~\ref{fig:rhos_vs_tperp}, the SF density $\rho_{\rm sf}$ is plotted against $t_\perp$ for $L=32,~48,~64,~96,~128,~192,~256,~512$, and normalized by its $t_\perp=0$ value which is simply, in the thermodynamic limit~\cite{Laflorencie01} $
\rho_{\rm sf}^{\rm 1D}={1}/{\pi}$. We observe in Fig.~\ref{fig:rhos_vs_tperp} a slow decay of the SF density when the interchain hopping is increased. Even for the largest system considered here ($1024$ sites), the SF response is still very important for $t_\perp <t/2$. While for large enough values of $t_\perp\simeq t$, the gapped nature of the system is obvious, with a vanishing SF density, it appears quite difficult to predict the existence of any critical coupling $t_\perp^c$ from the actual behavior of $\rho_{\rm sf}(t_\perp)$, as displayed in Fig.~\ref{fig:rhos_vs_tperp}.

\begin{figure}
\includegraphics[width=\columnwidth,clip]{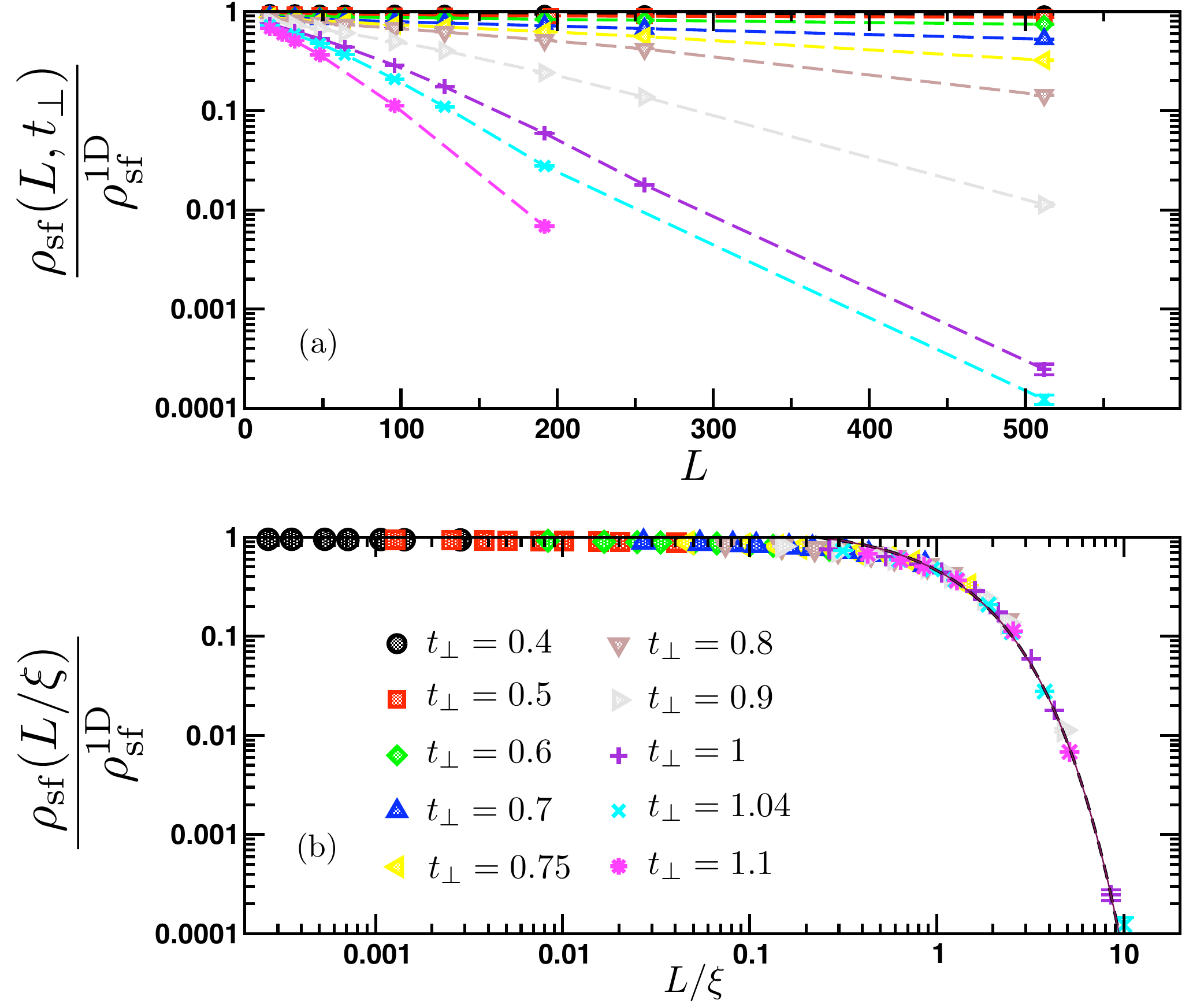}
\caption{\label{fig:scaling_stiff_ladder}
(Color online) (a) SF density plotted versus the linear size $L$ for various values of $t_\perp$, as indicated below with different symbols. The linear-log plot clearly shows an exponential decay with $L$. (b) Collapse of the above data obtained by rescaling the x-axis $L\to L/\xi$ in order to get all points on a single curve. The full line is $\propto \exp(-L/\xi)$. The obtained correlation length $\xi(t_\perp)$ are displayed in Fig.~\ref{fig:xi}.}
\end{figure}

\subsubsection{Scaling analysis}
We therefore use another strategy to extract the critical properties. As visible in Fig.~\ref{fig:scaling_stiff_ladder}(a) for various values of $t_\perp$, the SF density decay exponentially at large $L$, at least for not too small values of the coupling: $t_\perp>0.7t$, while it roughly remains constant for the sizes considered here when $t_\perp< 0.7 t$. We therefore identify two scaling regimes, depending on the scaling variable $x=L/\xi$, where $\xi$ is the correlation length of the RMI state:
\be
\label{eq:scaling_rhosf}
\rho_{\rm sf}/\rho_{\rm sf}^{\rm 1D}\simeq\left\{
\begin{array}{lr}
1&{\rm{~if~}} x\ll 1\\
\exp(-x) &{\rm{~if~}} x\gg 1
\end{array}
\right.
\ee

Data of Fig.~\ref{fig:scaling_stiff_ladder}(a) have been analyzed by rescaling the x-axis $L \to L/\xi$ in order to collapse the data : using the best fit of the form $A \exp(-L/\xi)$, we have fixed $\xi=61$ for $t_\perp=t$ and adjusted all the other data to get the best collapse. Results are displayed in Fig~\ref{fig:scaling_stiff_ladder}(b) where the two scaling regimes are clearly visible. With this technique, we have been able to extract the value of $\xi$ as a function of $t_\perp$ down to $t_\perp=0.4t$. For smaller values, it was almost impossible to see any downward curvature. 
\subsection{Correlation length}
\begin{figure}
\includegraphics[width=\columnwidth,clip]{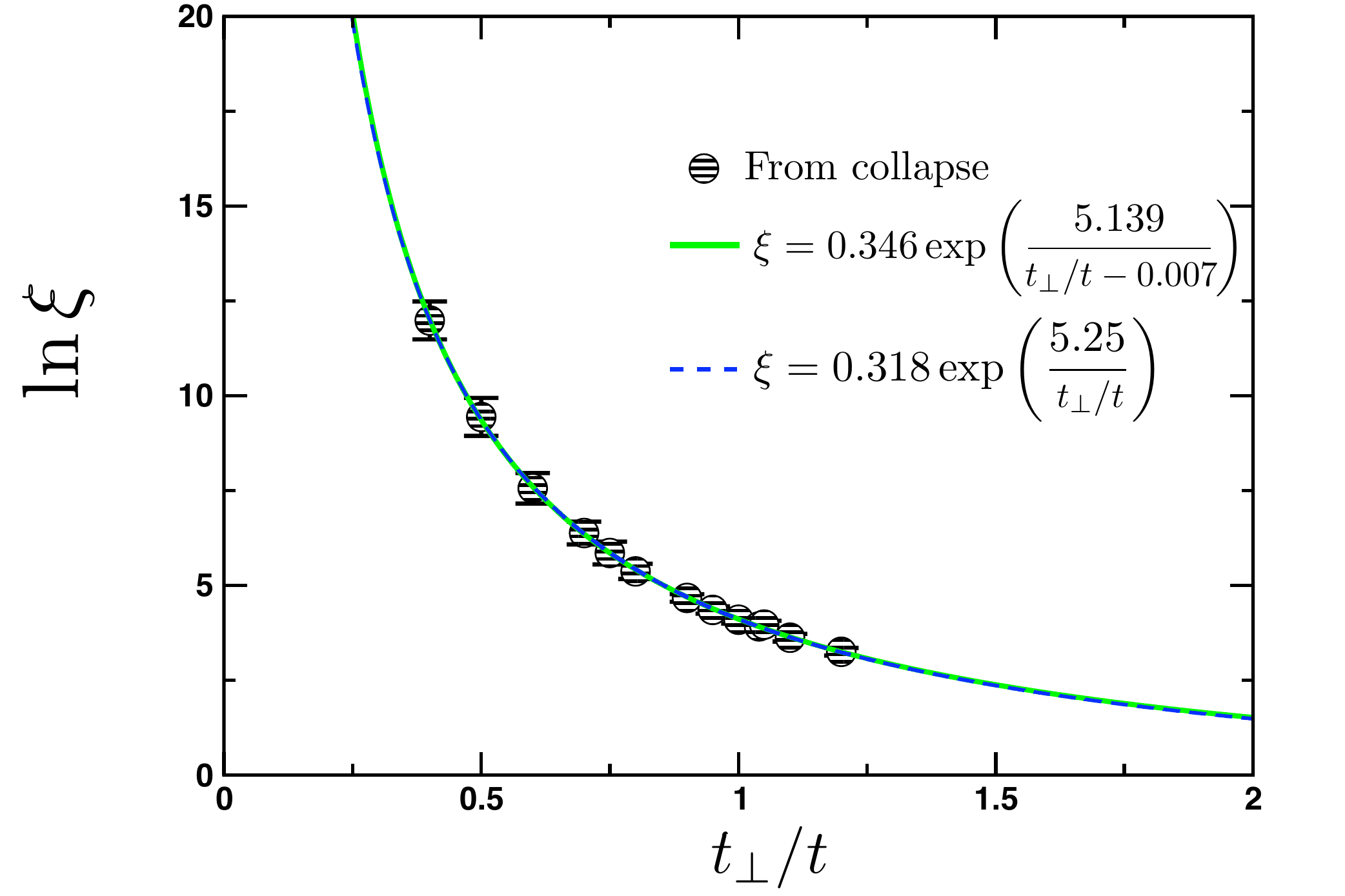}
\caption{\label{fig:xi}
(Color online) Correlation length of the bosonic ladder plotted versus the inter-chain hopping $t_\perp$. The points obtained from the data collapse of Fig.~\ref{fig:scaling_stiff_ladder} (black circles) are well-described by the exponential form displayed on the plot (green curve).}
\end{figure}

In order to convince oneself that the data collapse technique gives rather good estimates of the correlation length, we give in Appendix~\ref{app:collapse} a benchmark of the method in the case of the integrable $t-V$ model. Now we show how the correlation length of the bosonic ladder, extracted in the data collapse of Fig.~\ref{fig:scaling_stiff_ladder}, varies with $t_\perp$. We report the estimates in table~\ref{tab:xi}, and plot $\ln \xi$ against $t_\perp$ in Fig.~\ref{fig:xi} where two fitting functions of the form $\xi_0\exp\left[\text{a}t/\left({t_\perp^{\vphantom{c}}-t_\perp^c}\right)\right]$
are displayed. When leaving free the three parameters $\xi_0$, a, and $t_\perp^c$, we obtain for the best fit (green curve in Fig.~\ref{fig:xi}) a very small critical hopping $t_\perp^c=0.007t$, obviously compatible with the RG result $t_\perp^c=0$. When forcing  $t_\perp^c=0$ in the fitting process, we also obtain a very good agreement (dashed blue curve), which gives a very good support to the RG result
\be
\xi=\xi_0\exp\left({\text{a}}\frac{t}{t_\perp}\right).
\ee

Note that we can compare $\xi$ with the inverse gap $1/\Delta_s$ extracted from DMRG simulations (Eq.~(\ref{eq:gap}) and Fig.~\ref{fig:gap}). They both agree very well, showing in particular a rather large value of the parameter $\text{a}\simeq 5.2(1)$. As already emphasized, the physics of short-range correlations -- the usual hallmark of quantum spin ladders --  is not realized here, except in the limit of large interchain hopping $t_{\perp}\gg t$. It is indeed quite instructive to compare the results of $\xi$ for the bosonic ladder model with the correlation length of SU(2) symmetric Heisenberg ladders (table~\ref{tab:xi}), known to display very short-range correlations, thus providing the useful picture of short-range resonating valence bonds~\cite{White94}. In the bosonic ladder case, the RMI state has correlations along the legs which extend over a very large scale $\xi$, even for $t_\perp \sim t$, as one can read from table~\ref{tab:xi}. Practically speaking, for instance in an optical lattice setup made of at most several hundreds sites, two weakly coupled chains will behave as a gapless system (see the discussion in section~\ref{sec:exp}).
\begin{table}
\begin{tabular}{c|c|c}
${\underline{t_{\perp}}}$&$\xi$&$\xi$\\
$t$&{\rm{\scriptsize[bosonic ladder]}}&{\rm{\scriptsize[SU(2)~ladder]}}\\
\hline
1&$60(1)$&1.99\\
0.9&$106(2)$&2.28\\
0.8&$2.15(5)\times 10^2$&2.65\\
0.7&$5.9(1)\times 10^2$&3.12\\
0.6&$1.9(1)\times 10^3$&3.76\\
0.5&$1.2(1)\times 10^4$&4.64\\
0.4&$1.6(1)\times 10^5$&5.93\\
\hline
\end{tabular}
\caption{\label{tab:xi}QMC estimates for the correlation length $\xi$ of the half-filled RMI. These data are also plotted against $t_\perp$ in Fig.~\ref{fig:xi}. For comparison, the third column shows the correlation length of the corresponding rung singlet state of SU(2) Heisenberg ladders [model Eq.~\eqref{XXZ_ladder_hamiltonian}] with $J_\perp=J^z_\perp=t_\perp$ and $J^z=J=t$ (see also Fig.~\ref{fig:gap_comp}).}
\end{table}

\section{Luttinger liquid behavior}
\label{sec:LL}

We now turn to the case of incommensurate fillings, and study the behavior of the Luttinger parameter $K_s$
of the symmetric gapless mode which governs all correlators. It has a
non-trivial dependence on density and interchain hopping which
requires non-perturbative numerical estimates : two complementary
approaches are used from DMRG and QMC calculations. Analytical results for several limiting cases are also presented.
A first account of this behavior was given in Ref.~\onlinecite{Hikihara2001}. We  improve the numerical accuracy of the
calculation and we present a detailed discussion of the physical origin of the curves.

\subsection{Luttinger parameter from the correlation functions: DMRG results}

\begin{figure}
\centering
\includegraphics[width=\columnwidth,clip]{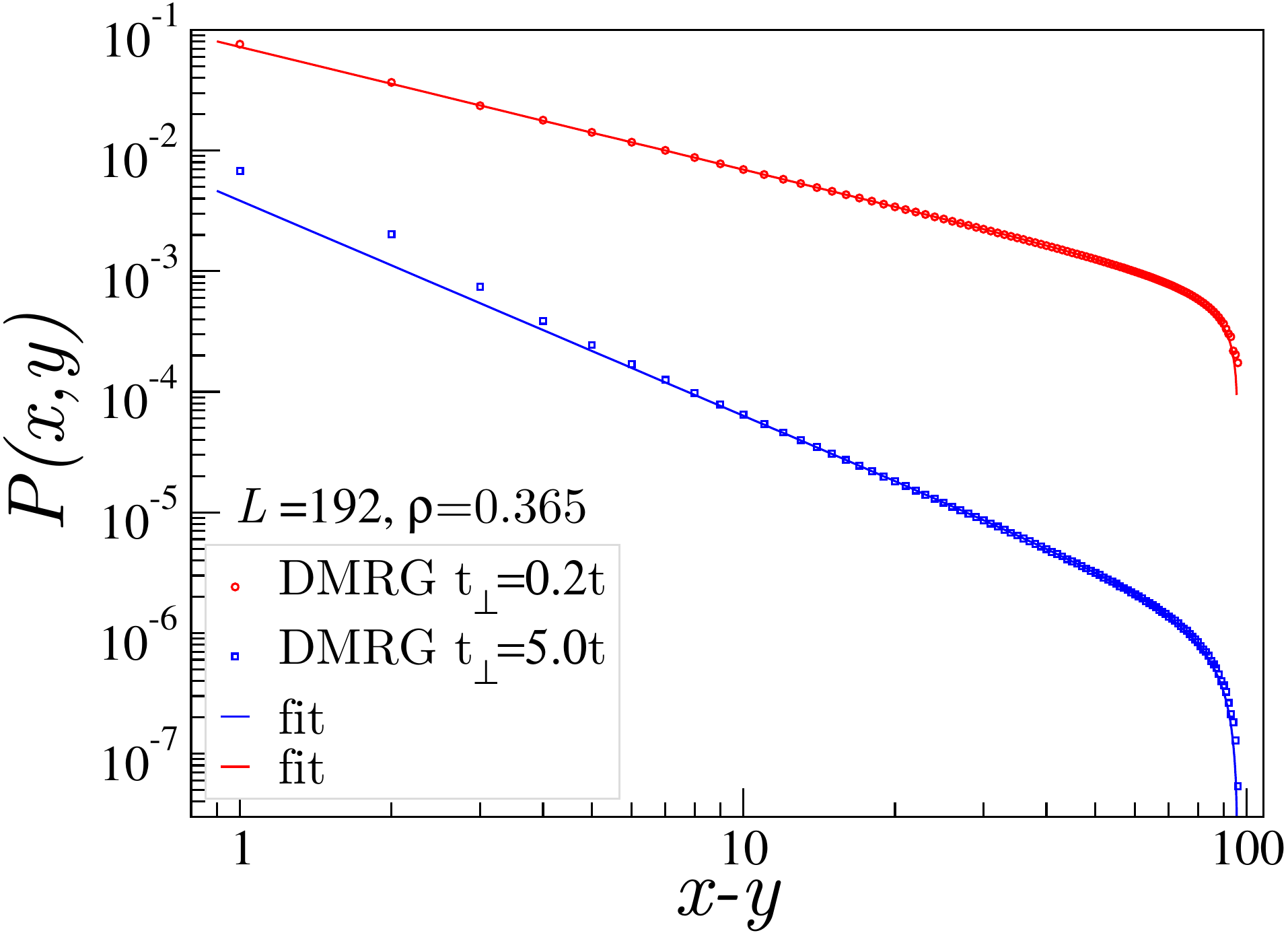}
\caption{\label{fig:Correlations_DMRG}(Color online) Typical rung-pairing
  correlation function $P(x)$ computed with DMRG and fitted by
  Eq.~\eqref{eq:fitcorr}. The effect of the interchain hopping
  $t_\perp$ is strong on both the amplitude and the exponent.}
\end{figure}

In this section, we describe how $K_s$ is extracted from the
computation of correlation functions using DMRG with open boundary
conditions (OBC). As $K_s$ governs the decay of all algebraic
correlators, the idea is to find a suitable order parameter which
provides a direct access to $K_s$. The most natural one would be the
bosonic propagator along a chain
$\langle{b_{1,j+x}^{\dag}b_{1,j}^{\dagga}}\rangle$, which has a leading
algebraic decay at zero-momentum governed by the exponent $4/K_s$. However, numerical
calculations shows that the signal is spoiled by subleading
oscillating terms. 
A better choice is actually to look at the propagator of a pair of bosons
created on a rung, defining:
\begin{equation}
P(x,y) = \langle{b_{1,x}^{\dag}b_{2,x}^{\dag}b^{\dagga}_{1,y}b^{\dagga}_{2,y}}\rangle\;.
\end{equation}
This correlator has a zero-momentum leading term with a decay exponent
$1/K_s$ and do not display sizable subleading corrections as shown in
Fig.~\ref{fig:Correlations_DMRG}. The data are here calculated from
the middle to the edge of a system with $L=192$ and keeping 2000
states. With OBC, the wave-function naturally vanishes at the edges, 
causing a fall of the correlator when $x,y\simeq L$. In order to
get the best estimates for $K_s$, we have to take into account this
finite-size corrections. Thanks to the conformal field theory results
of Refs.~\onlinecite{Hikihara2001, Cazalilla2004} and which already proved to give
very good fits of DMRG data~\cite{Hikihara2001, Roux2008, Roux2009}, we use the
following fitting function for a superfluid-like correlator:
\begin{equation}
P(x,y) = A \left[\frac{\sqrt{d(2x|2L)d(2y|2L)}}{d(x+y|2L)d(x-y|2L)}\right]^{1/K_s}\;,
\label{eq:fitcorr}
\end{equation}
in which we take $y=L/2$. Here, $d(x|L) =
\frac{L}{\pi}\sin\left(\frac{\pi x}{L}\right)$ is the cord
function~\footnote{Taking $L+1$ instead of $L$ in the formulas is
  actually more correct on finite system as the wave-function vanishes
  for $j=0,L+1$ but in the case under study, this leads to
  insignificant corrections.}. Typical fits are displayed in
Fig.~\ref{fig:Correlations_DMRG}, from which a rather accurate decay
exponent can be extracted.

\begin{figure}
\centering
\includegraphics[width=\columnwidth,clip]{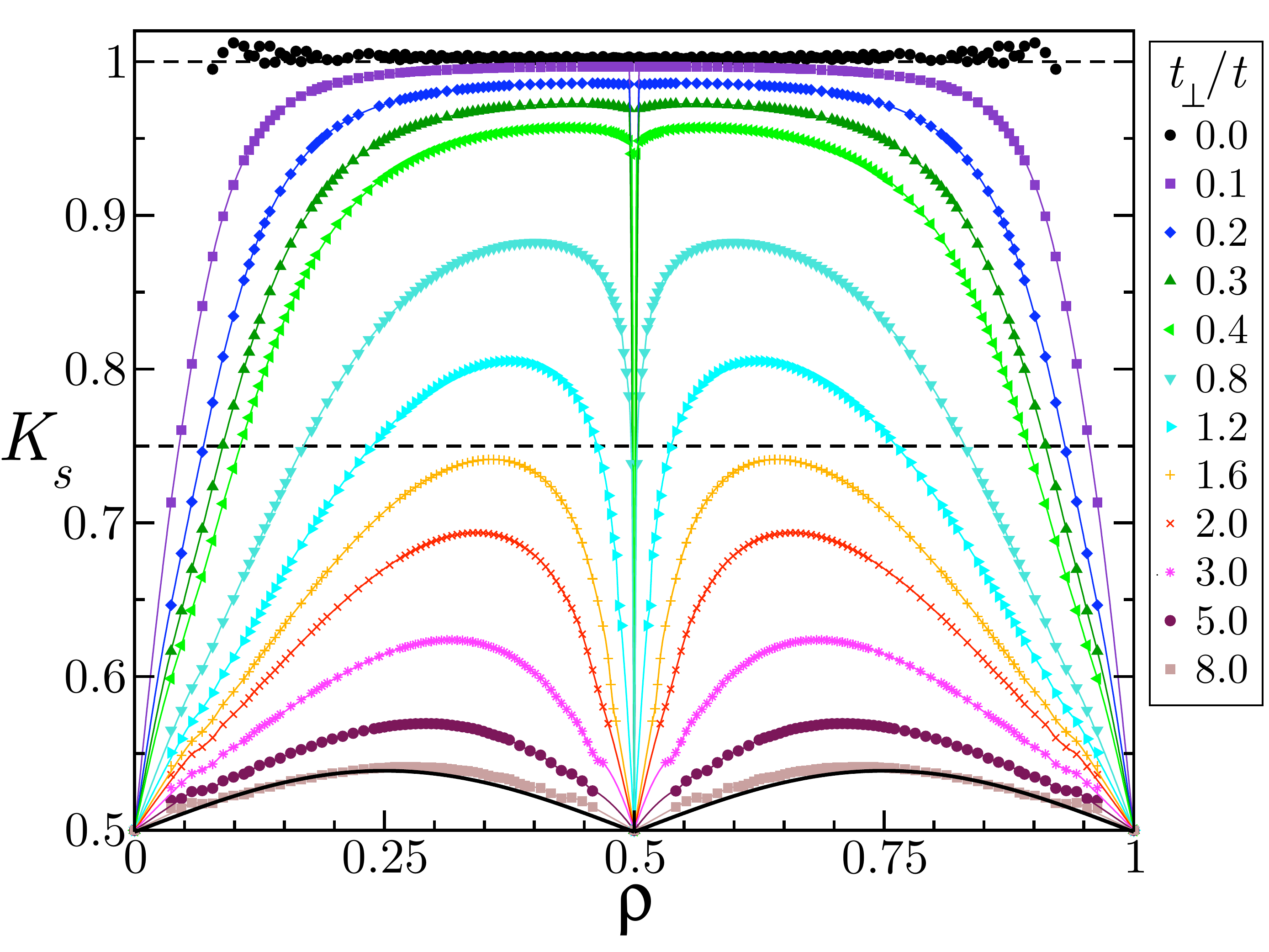}
\caption{\label{fig:K_DMRG}(Color online) DMRG results for the Luttinger parameter
  $K_s$ plotted versus the density $\rho$. Various symbols represent
  different interchain hopping strengths $t_\perp/t$. The full black line is the perturbative results Eq.~\eqref{eq:K_SRC} for $t_\perp/t=8$, as derived in Sec.~\ref{sec:K_SRC}.}
\end{figure}

Using this procedure, we mapped out the behavior of $K_s$ as a
function of the density $\rho$ for various interchain hopping
$t_\perp$. The behavior is plotted in Fig.~\ref{fig:K_DMRG} (in a
symmetrical way because of particle-hole symmetry) and shows several
striking features. $K_s$ is strongly affected, as expected, by the
vicinity of the commensurate-incommensurate (C-IC) transition occurring at
$\rho=1/2$. Between 0 and 0.5, a strong asymmetry is found,
which is surprisingly qualitatively very similar to the behavior of
the Luttinger parameter of the charge mode in the 1D fermionic Hubbard
model~\cite{Schulz1990}, and for which we will give supporting
arguments. The limiting value of $K_s$ is naturally one in the limit
of uncoupled chains~\footnote{Although a simulation of two uncoupled
  chains makes little sense, we show on Fig.~\ref{fig:K_DMRG} that the
  obtained values for $K_s$ are surprisingly rather good. A finite
  $t_\perp$ leads to better convergence of course and the
  corresponding values are much more reliable.}. Before embarking in a more careful analysis, we may
notice that the C-IC scenario tells that the expected value of $K_s$
approaching the commensurate lines by varying the density is half that
of the critical $K_s$ obtained by varying the coupling which does yield $K_s=1/2$ in the vicinity of $\rho=1/2$.

\subsection{QMC estimate of the Luttinger parameter}

\begin{figure}
\includegraphics[width=7.5cm,clip]{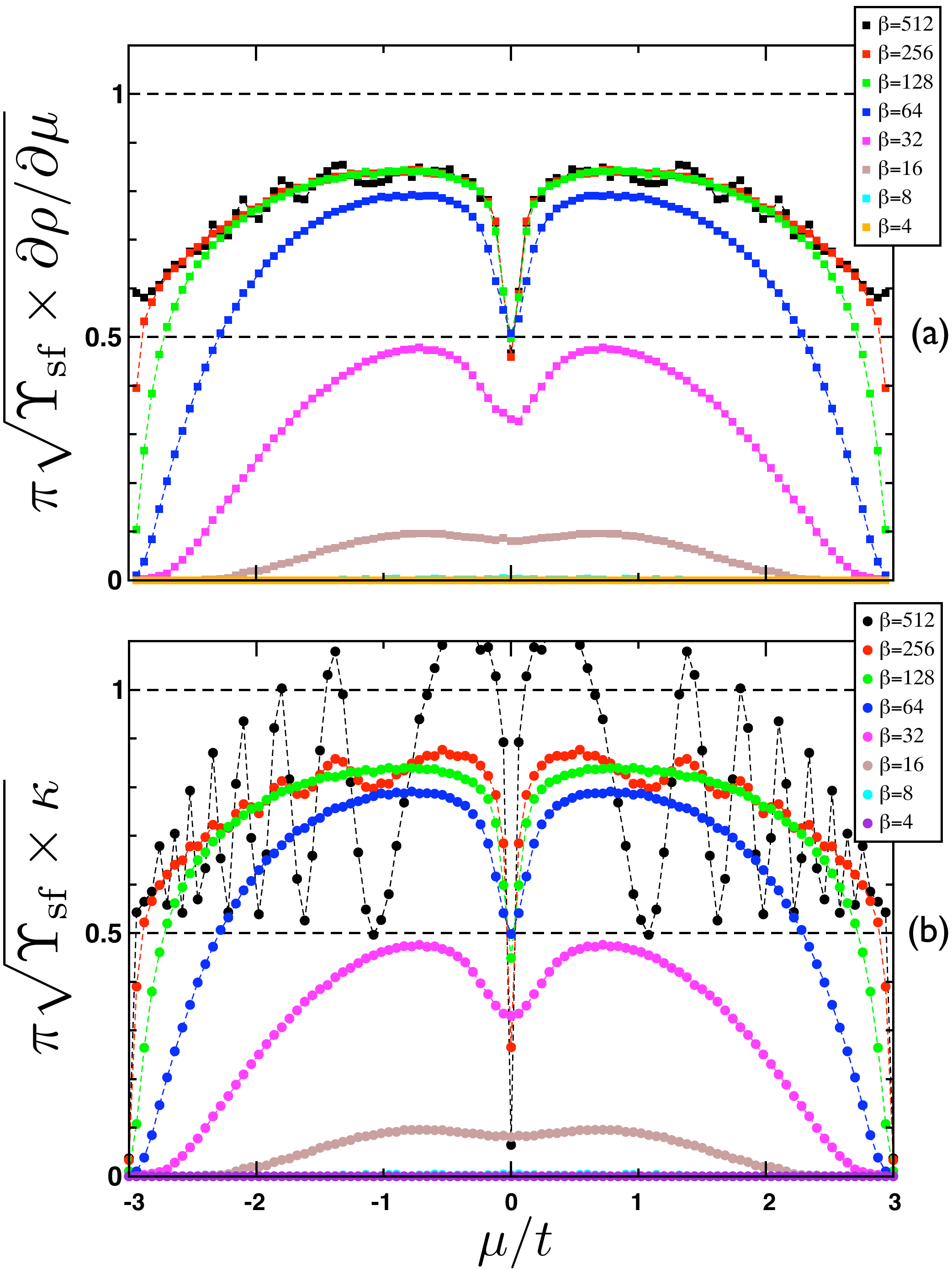}
\caption{\label{fig:Keff_tperp1}
(Color online) Effective LL parameter $K_{\rm eff}(T)=\pi\sqrt{\kappa(T)\Upsilon_{\rm sf}(T)}$ computed with SSE simulations on a $2\times 64$ ladder with $t_{\perp}=t=1$ and for various temperatures, as indicated on the plot. The compressibility is computed either through the numerical derivative $\partial \rho/\partial \mu$ (a) or directly using the SSE estimate Eq.~(\ref{eq:kappa}) (b). Note that large oscillations visible at the lowest temperatures disappear when approaching the transition points $\mu_c=\pm3$.}
\end{figure}

The Luttinger parameter of the gapless (symmetric) mode can also be computed quite efficiently using QMC. While the Green's functions are in principle accessible with SSE~\cite{Dorneich01}, a precise determination of long distance off-diagonal correlators of the form $\langle b_{j}^{\dagger}b_{j+r}^{\dagga}\rangle$ is not an easy task. An interesting alternative using QMC is to take advantage of the "hydrodynamic relation" of the LL where phase and density fluctuations are related through
\be
K_s=\pi{\sqrt{\kappa\Upsilon_{\rm sf}}}.
\label{eq:KQMC}
\ee
%

\begin{figure}
\includegraphics[width=8cm,clip]{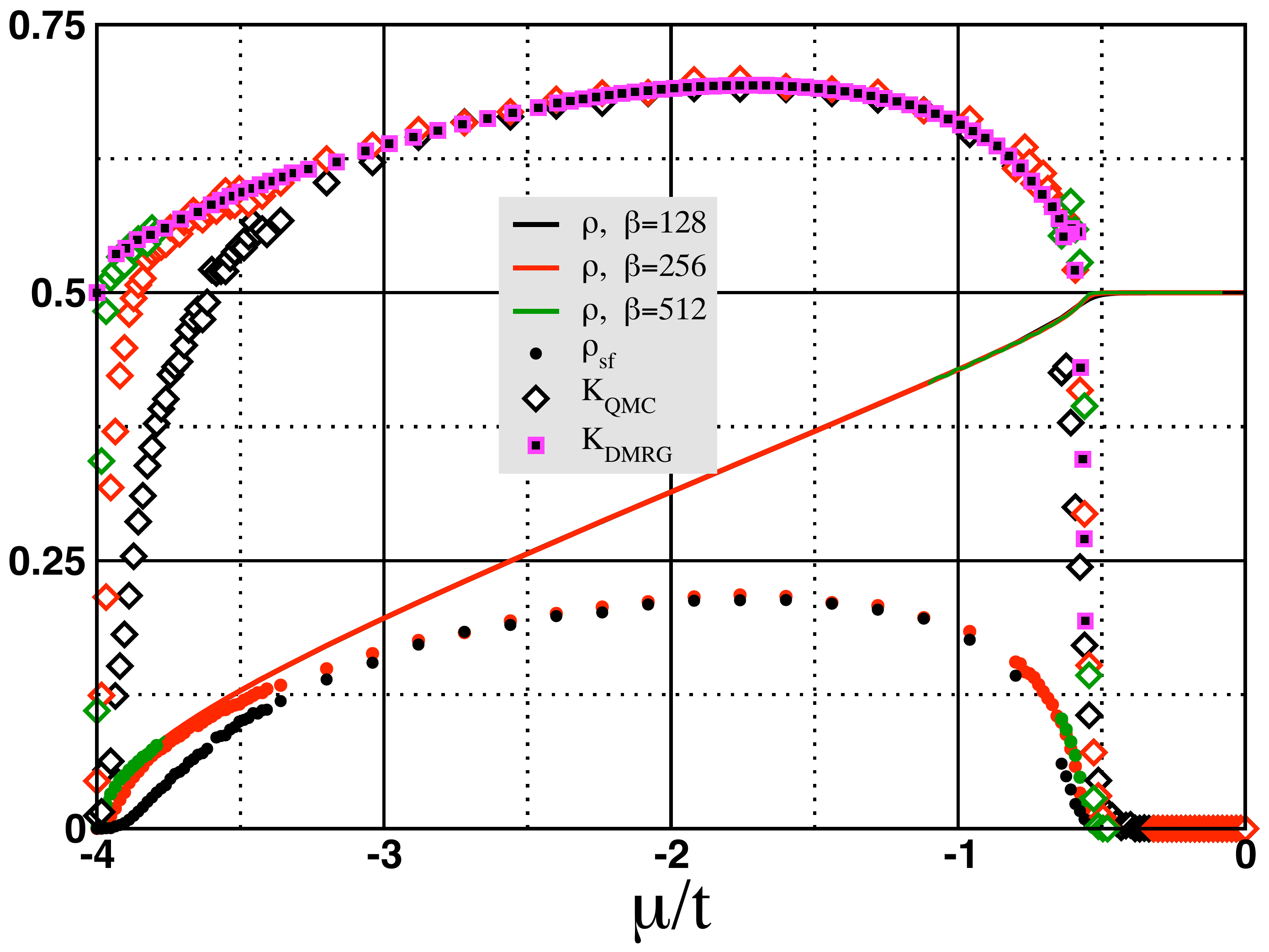}
\caption{\label{fig:tperp2}
(Color online) QMC results obtained for $t_{\perp}=2t$ and $L=64$ at various temperatures. The full lines show the total particle density $\rho$ whereas the SF part $\rho_{\rm sf}$ is shown by small circles for the three values of $\beta$ using the same color as for $\rho$. The Luttinger parameter $K_s$, extracted using Eq.~(\ref{eq:KQMC}), is plotted also for the three considered temperatures (colored diamonds), and compared with DMRG estimate (magenta squares).}
\end{figure}

As discussed above, it is straightforward to compute the superfluid stiffness $\Upsilon_{\rm sf}$ and the compressibility $\kappa$. We therefore compute the effective LL parameter $K_{\rm eff}(T)=\pi\sqrt{\kappa(T)\Upsilon_{\rm sf}(T)}$.
Finite temperature results are shown for a $2\times 64$ ladder with $t_\perp=t$ in Fig.~\ref{fig:Keff_tperp1} where one sees three different regimes. (i) $K_{\rm eff}(T)=0$ at high temperature because the SF stiffness is zero. (ii) When $T$ decreases $K_{\rm eff}(T)$ increases and saturates as soon as $T$ becomes of the order of the finite size gap $\Delta(L)$. (iii) For $T$ below $\Delta(L)$, oscillations in the compressibility $\kappa$ start to appear, thus signaling the appearance of small magnetization steps in the GS of the finite length system. This effect naturally induces oscillations of $K_{\rm eff}(T)$. Note that these oscillations are more pronounced when $\kappa$ is directly computed through Eq.~(\ref{eq:kappa}) whereas taking the numerical derivative $\partial \rho/\partial \mu$ has a smoothing effect, as visible in Fig.~\ref{fig:Keff_tperp1} (a-b).
It is however interesting to note that close to the diluted limit, where the level spacing at the bottom of the cosine band becomes much smaller $\Delta(L)\sim L^{-z}$ with $z=2$, these oscillating features tend to disappear, even at the lowest temperature considered here. It appears very clearly here that the Luttinger parameter tends to 0.5 in this diluted limit, as well as in the other limit, when the RMI is approached. Below, we give analytical support to such a limit. 
In between, the incommensurate regime is characterized by a $\mu-$dependent Luttinger exponent $K_s(\mu)$ which takes non-universal values in the range $(0.5 - 1)$. These results are in very good agreement with DMRG estimates discussed above,
%
%
as shown in Fig.~\ref{fig:tperp2} where QMC together with DMRG results are displayed for $t_\perp/t=2$. Again one sees that a very large inverse temperature $\beta$ is necessary to capture the GS properties close to the diluted limit. In particular this is quite visible in Fig.~\ref{fig:tperp2} for the SF density $\rho_{\rm sf}$ which displays important finite temperature effects in this diluted regime.

We now give some physical interpretation which supports the numerically observed Luttinger parameter $K_s$ in various limiting cases.
\subsection{Limiting cases: analytical results}
\subsubsection{Dilute Bose gas limit}
\label{sec:DBG}
\begin{figure}
\includegraphics[width=\columnwidth,clip]{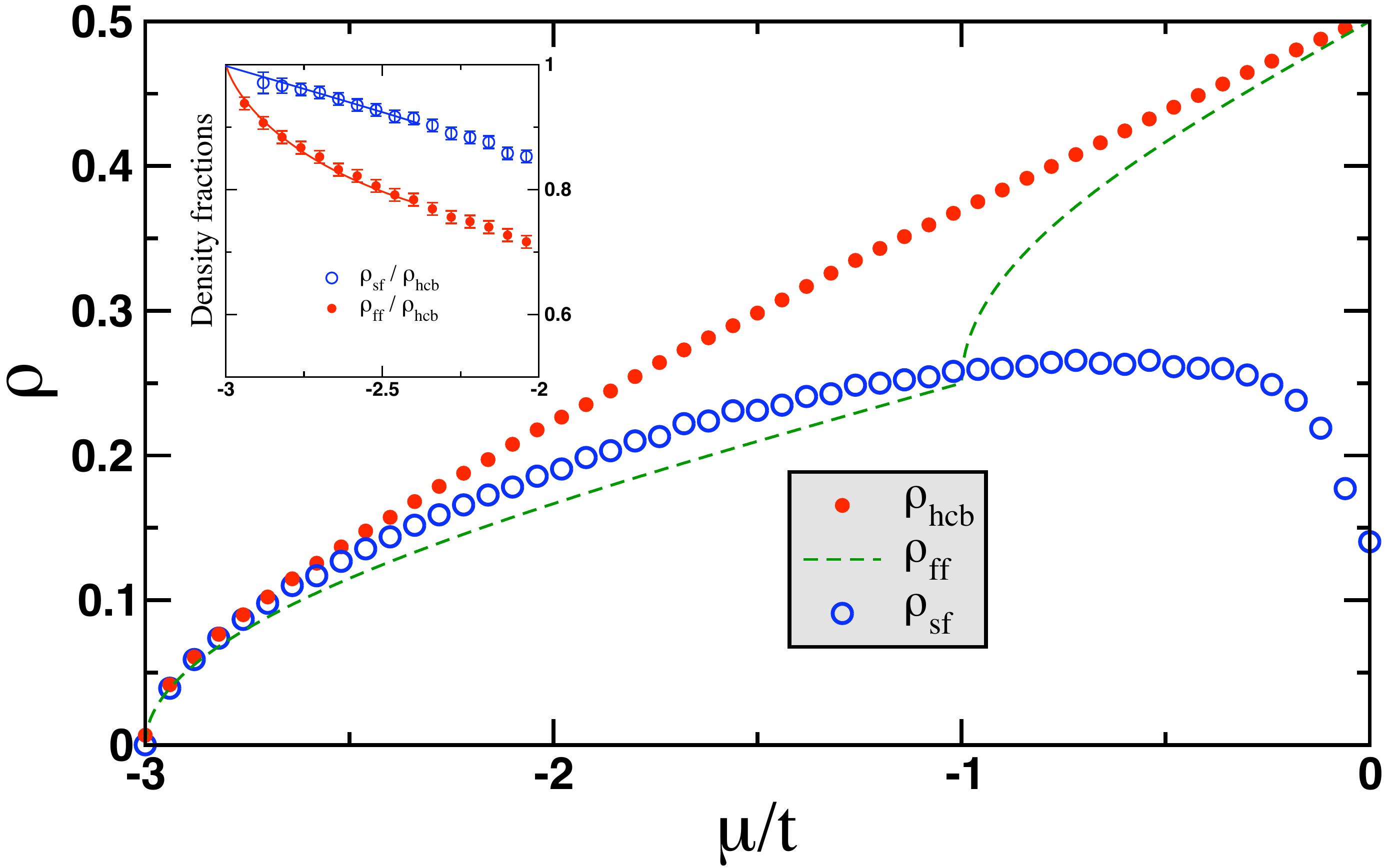}
\caption{\label{fig:QMCdiluted}
(Color online) Various densities plotted vs $\mu/t$ for $t_\perp=t$: hard-core bosons (QMC, red circles), free-fermions (exact, dashed line), and the SF density for the hard-core bosons (QMC, open circles). QMC results obtained for $2\times 64$ ladders at $\beta/t=512$. Inset: SF fraction $\rho_{\rm sf}/\rho_{\rm hcb}$ (open blue circles) and fermion-boson ratio $\rho_{\rm ff}/\rho_{\rm hcb}$ (red circles) shown close to the critical point $\mu_c=-3t$. The lines are guides to the eyes.
}\end{figure}

In the extremely diluted limit, the bosonic density grows with the chemical potential $\mu$ as the density of free fermions would do upon filling up the bottom of the cosine $0$-band Eq.~(\ref{eq:0}). This is apparent in the srong transverse hopping limit, where the Hamiltonian is actually mapped onto a band of spinless fermions in equation \eqref{eq:Hefffermions}. At low densities, interactions and correlated hopping can be neglected, and fermions are indeed free. In the weak coupling limit, a look at bosonic propagators helps to draw a similar picture. Considering symmetric $b_{j,+} = (b_{j,1}+b_{j,2})/\sqrt{2}$, and antisymmetric bosons $b_{j,-} = (b_{j,1}-b_{j,2})/\sqrt{2}$, we have:
\beq
\label{eq:propsym}
\hspace{-0.3cm}\langle b_+^\dagger(x)b_+^\dagga(0)\rangle \sim \left( \frac{\alpha}{x} \right)^{\frac{1}{4K_s}}\langle \cos\left(\frac{\theta_a(x)}{\sqrt{2}}\right)\cos\left(\frac{\theta_a(0)}{\sqrt{2}}\right)\rangle, 
\eeq
and
\beq
\langle b_-^\dagger(x)b_-^\dagga(0)\rangle \sim \left( \frac{\alpha}{x} \right)^{\frac{1}{4K_s}}\langle \sin\left(\frac{\theta_a(x)}{\sqrt{2}}\right)\sin\left(\frac{\theta_a(0)}{\sqrt{2}}\right)\rangle.
\eeq
Since interchain hopping is a relevant perturbation and the field $\theta_a$ is gapped, the average of cosines decay exponetially to 1, over a length scale $\xi_a$, while the average of sines decay exponentially to 0 over the same length scale. In the next paragraph, we discuss in detail the behavior of the important length scale $\xi_a$ with $t_\perp$ and the filling $\rho$. Here it is enough to realize that at very low fillings, only one gapless mode remains, consisting of a Luttinger liquid of symmetric bosons. In Fig.~\ref{fig:QMCdiluted}, we computed both the filling of hard-core bosons and free fermions on a two-leg ladder with $t=t_\perp$. It appears that they become identical in the extremely diluted limit, confirming both our strong and weak coupling analysis. This leads to the following square-root behavior for the density
\be
\rho(\mu)=\frac{1}{2\pi}\sqrt{\frac{\mu-\mu_c}{t}},
\ee
which yields a compressibility
\be
\kappa(\mu)=\frac{1}{4\pi\sqrt{t(\mu-\mu_c)}}.
\ee
Also, for very diluted systems, it is natural to expect that the SF fraction will tend to unity~\cite{Leggett98} : $\rho_{\rm sf}/\rho\to 1$. Again this is exemplified in Fig.~\ref{fig:QMCdiluted} where $\rho_{\rm sf}/\rho_{\rm hcb}\to 1$ as $\rho_{\rm hcb} \to 0$. Therefore, in such a limit the stiffness $\Upsilon_{\rm sf}=2t\rho_{\rm sf}$ is simply given by
\be
\Upsilon_{\rm sf}=\frac{t}{\pi}\sqrt{\frac{\mu-\mu_c}{t}}
\ee
which, using the hydrodynamic relation for the Luttinger parameter Eq.~(\ref{eq:KQMC}) simply becomes
\be
K_s=\frac{1}{2}.
\ee
%
\noindent This is a striking difference between free fermions and hard-core bosons on a two-leg ladder: the former have a single-mode LL with $K=1$ even in the vicinity of $\rho=0,1$. Note that symmetric bosons indeed behave as 1D hard-core bosons at very low fillings since their propagator now reads $\langle b_+^\dagger(x)b_+^\dagga(0)\rangle \sim x^{-1/2}$. From Fig.~\ref{fig:QMCdiluted} we also notice that as the density increases, the following inequality $\rho_{\rm hcb}> \rho_{\rm sf} > \rho_{\rm ff}$ is verified, implying that $K_s> 1/2$ as the system is filled up. 
\subsubsection{Doped RMI: strong rung coupling}
\label{sec:K_SRC}
Starting from the effective spinless fermions Hamiltonian derived in Sec.~\ref{sec:EffMod} to second order in $t/t_\perp$, equation \eqref{eq:Hefffermions}, we deal with a 1D chain for which it is possible to compute both the Luttinger parameter $\tilde{K}$ and the renormalized velocity $\tilde{v}$ using the compressibility and the phase stiffness. Indeed one can easily compute the GS energy of the effective Hamiltonian, to first order in $t/t_\perp$ and its response to a change in the number of particles or a twist in the boundary conditions (see for instance Ref.~ \onlinecite{Cazalilla04} as well as appendix \ref{app:K}). We find:
\begin{equation}
\tilde{K} = 1 + \frac{2t}{t_\perp}\frac{\sin(\pi\tilde{\rho})}{\pi},
\end{equation}
with $\tilde{\rho} = 2(\rho-0.5)$ the density of fermions, that is, the density of doubly occupied rungs. We now turn to the question of the relation between $\tilde{K}$ and $K_s$ in the original system. As argued earlier, intuition suggests that the fermions in the effective model are related to the symmetric bosons of the original ladder, $b_+=(b_1 + b_2)/\sqrt{2}$, and, from the related propagator \eqref{eq:propsym}, we saw that at large enough distances, $x \gg \xi_a$, these particles actually behave as a Luttinger liquid. In the strong coupling regime, using the mappings of Appendix \ref{app:effmodel}, we calculate the propagator to be $\langle b_+^\dagger(x)b_+^\dagga(0)\rangle \sim x^{-1/(2\tilde{K})}.$ We can readily make the identification $\tilde{K}=2K_s$ in the strong coupling regime and obtain the following perturbative expression:
\begin{equation}
K_s = \frac{1}{2} + \frac{t}{t_\perp}\frac{\sin[2\pi(\rho-0.5)]}{\pi}\,.
\label{eq:K_SRC}
\end{equation}
The latter expression seems already a good approximation for $t_\perp = 8 t$, as visible in Fig.~\ref{fig:K_DMRG}. Note that a similar calculation was first carried out by Cazalilla\cite{Cazalilla04} in a different context. There, the author computed the Luttinger parameter of the Bose-Hubbard model in the limit of strong on-site interaction by mapping the problem onto a chain of spinless fermions with attractive nearest-neighbor interactions and next-nearest neighbor correlated hopping. A parallel can be drawn in our case. One could argue that the symmetric bosons are effectively soft-core bosons, since one can put two of them on the same site. However it is important to note that one cannot put more than two, and that a perturbative calculation on the Bose-Hubbard model for fillings greater than one would lead to a different effective Hamiltonian. 

Naturally, by doping the rung-Mott insulator with holes, one recovers a symmetric expression for $K_s$ when $\rho<0.5$:
\begin{equation}
K_s = \frac{1}{2} + \frac{t}{t_\perp}\frac{\sin[2\pi(0.5-\rho)]}{\pi}\,.
\end{equation}
This being said, we now have a quantitative description of the behavior of $K_s$ for large $t_\perp/t$. 
\subsubsection{Weak coupling limit and asymmetry of $K_s$}
We now provide a simple argument taken from the RG analysis to explain the qualitative behavior 
of $K_s$ at low $t_\perp/t$, in particular the large asymmetry about quarter fillings. For $t_\perp < t$, the length $\xi_a$ we introduced earlier can be computed from the RG:
\begin{equation}
\xi_a \sim (t_\perp /v)^{-2K/(4K-1)},
\label{eq:xia}
\end{equation}
where $v=2t\sin(\pi\rho )$ is the sound velocity along the chains. Such a length scale can be seen as the typical distance above which the superfluid phase fields of the two chains lock together. It is interesting to look at the ratio $\xi_a/d$, where $d$ is the mean "relevant distance" between particles.
By relevant, we mean the distance between bosons: $d\sim(2\rho)^{-1}$ for $0<\rho<1/4$ and between holes $d\sim(1-2\rho)^{-1}$ for $1/4<\rho<1/2$. There are two regimes of interest. (i) When $\xi_a\ll d$, the system is effectively 1D since the transverse phase fluctuations appear to decay over a distance much smaller than the inter-particle distance $d$: in such a regime, we expect $K_s\sim 1/2$. (ii) In the opposite limit $\xi_a\gg d$, the system will behave as two weakly interacting SF for which one expects $K_s\sim 1$. 

\begin{figure}
\begin{center}
\includegraphics[width=\columnwidth]{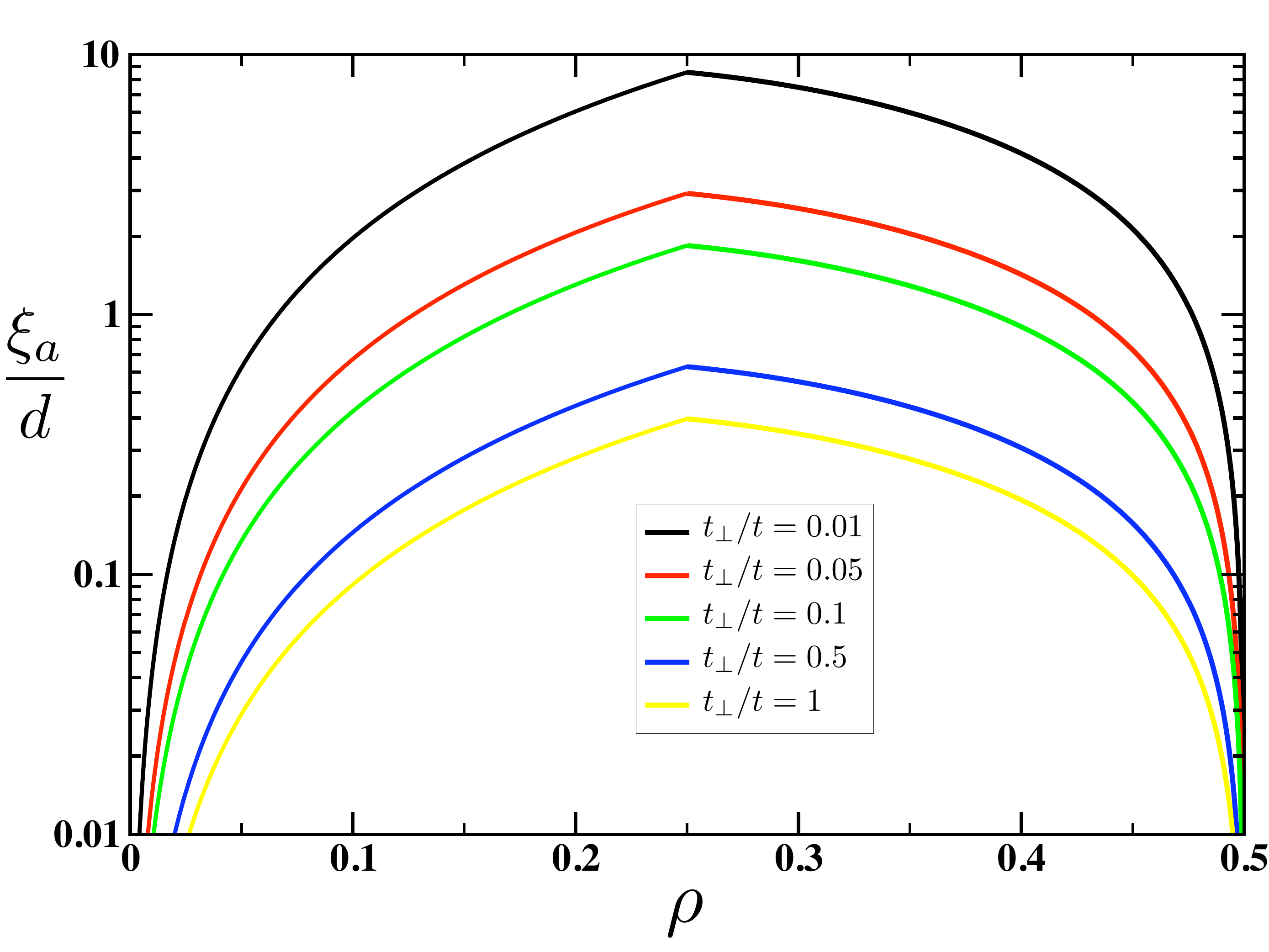}
\caption{\label{fig:xiperp}(Color online) Ratio $\xi_a/d$ as a function of the density $\rho$ for different values of $t_\perp/t$.}\end{center}
\end{figure}

One can try to get a slightly more quantitative view on this,
replacing $K=1$ in Eq.~\eqref{eq:xia}, we get 
\be
\frac{\xi_a}{d}={\rm min}\left[(2\rho),(1-2\rho)\right] \times \left(2\sin(\pi\rho)\frac{t}{t_\perp}\right)^{2/3}.
\ee
Plotted in Fig.~\ref{fig:xiperp}, this ratio
displays an asymmetric behavior about $\rho=1/4$ which is connected with the asymmetric behavior of $K_s$ at small $t_\perp/t$ (Fig.~\ref{fig:K_DMRG}). In particular, the much sharper vanishing of $\xi_a/d$ close to half-filling than in the diluted limit explains the sharp decreasing to $K_s=1/2$ at half-filling for small values of $t_\perp/t$ whereas at zero (or unit) filling $K_s$ goes to $1/2$ much slower (see Fig.~\ref{fig:K_DMRG}). One also sees that when $t_\perp/t$ increases, $\xi_a/d$ becomes very small for all fillings. This clearly signals that the effective 1D model obtained above in the other limit ($t_\perp/t\gg 1$) becomes a good description and that the asymmetry tends to disappear, as already noticed.

\subsubsection{Critical velocity}
%
\begin{figure}
\includegraphics[width=\columnwidth,clip]{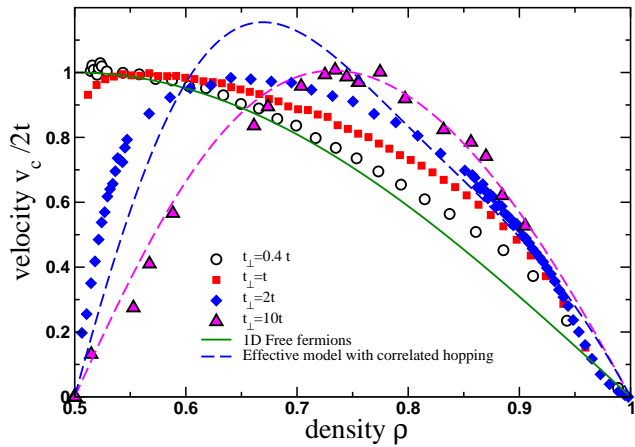}
\caption{\label{fig:vc}
(Color online) Velocity $v_c/2t$, extracted from QMC simulations using Eq.~(\ref{eq:vc}), plotted against the density $\rho$ for various values of the interchain hopping $t_\perp$, and compared with two limiting cases (see text): Eq.~(\ref{eq:vclimit}) in the weak coupling limit (green full line), and Eq.~\eqref{eq:vceff} for the strong coupling limit (blue dashed line).}\end{figure}

We finish this discussion with the analysis of the speed of sound $v$ discussed above. This velocity can be interpreted as the SF critical velocity $v_c$ associated to the linear branch in the excitation spectrum. Using the same hydrodynamic relations as above, we get for the velocity
\be
v_c=\sqrt{\frac{\Upsilon_{\rm sf}}{\kappa}}.
\label{eq:vc}
\ee
QMC results are shown in Fig.~\ref{fig:vc} for $v_c$ versus the density $\rho$ in various limiting cases $t_\perp=0.4 t,~t,~2t,~10t$. For a very small interchain hopping we expect the decoupled chains result to hold for $v_c$, that is:
\be
\label{eq:vclimit}
v_c = 2t\sin\left(\pi\rho\right) \ \ {\rm{~if~}} t_\perp\ll t,
\ee
whereas in the very strong rung coupling limit, the effective 1D model Eq.~(\ref{eq:Hefffermions}) with correlated hopping terms yields a different velocity of excitations as follows:
\beq
\label{eq:vceff}
v_c = 2t\left|\sin\left(2\pi\rho\right)\right|\left[1-\frac{4t}{t_\perp}(1-\rho)\cos(2\pi \rho)\right] \ \ {\rm{~if~}} t_\perp\gg t.
\eeq
The latter expression was obtained by following the calculation that led to the perturbative expression of $K_s$, equation \ref{eq:K_SRC}. We computed the compressibility and superfluid stiffness of the effective model and used the hydrodynamic relation \eqref{eq:vc}. The agreement between these two limiting cases and the QMC estimates for $v_c$ are quite good, especially for the two limits $t_\perp=0.4 t$ and $t_\perp=10t$. In between, for instance for $t_\perp=2t$, the shape of $v_c(\rho)$ is clearly non-universal, and asymmetric.\\

%
\begin{figure}
\includegraphics[width=\columnwidth,clip]{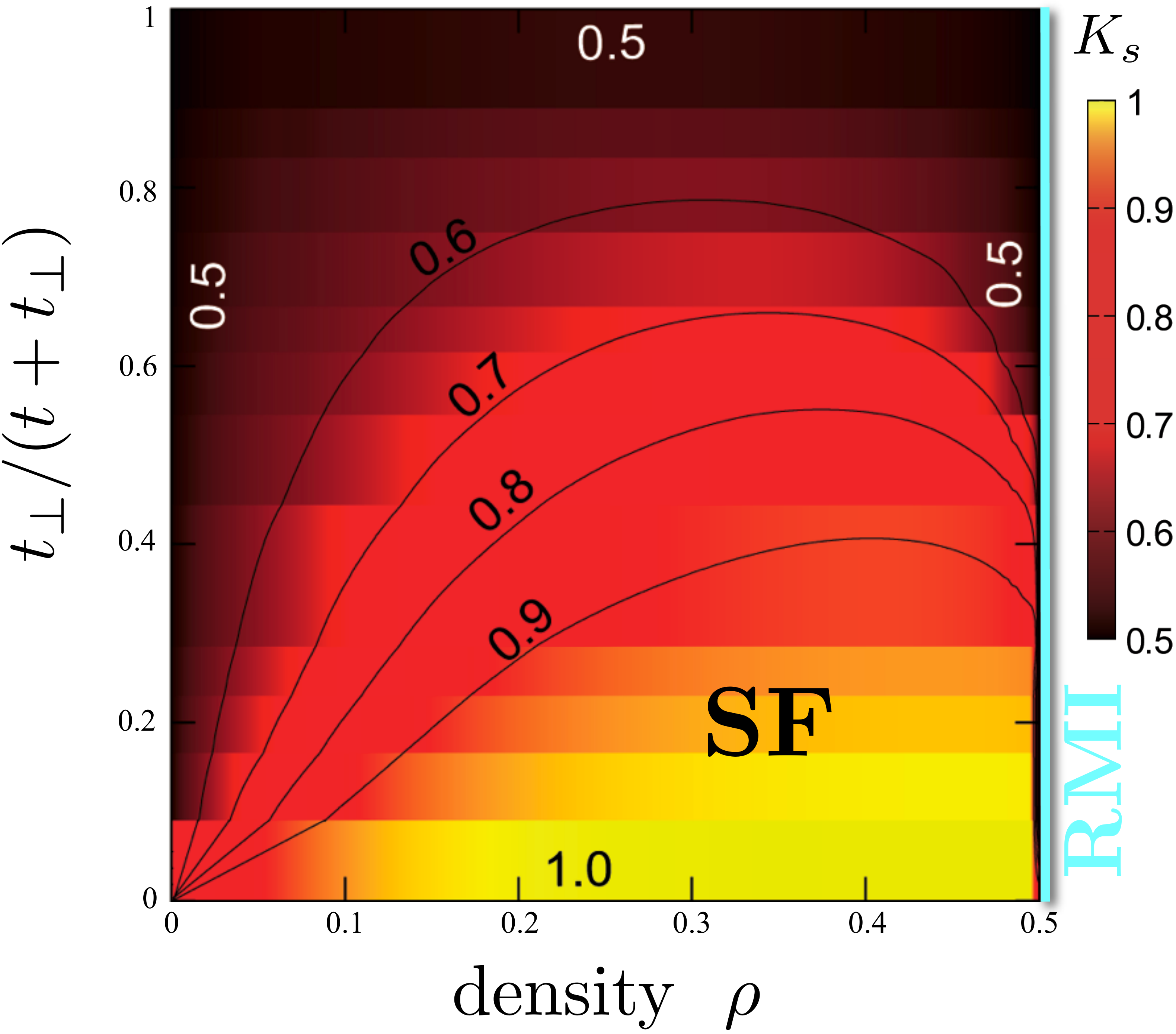}
\caption{\label{fig:map_rho}(Color online) Phase diagram for hard-core bosons on a two-leg ladder Eq.~\eqref{eq:model} in the plane density $\rho$ - interchain hopping $t_\perp/(t+t_\perp)$. The contour map for the Luttinger liquid parameter of the gapless symmetric mode $K_s$ has been obtained from DMRG simulations. At $\rho=1/2$ there is the gapped RMI (cyan line), and a SF phase for all other incommensurate fillings with $K_s\in[1/2,1]$.}
\end{figure}

\section{Discussions}
\label{sec:disc}
\subsection{Phase diagram}
Before discussing further two important implications of our results, namely the effect of weak disorder on the SF phase and the consequences for experiments, we present the general phase diagram of the hard-core bosons model on a two-leg ladder Eq.~\eqref{eq:model}. As visible in Fig.~\ref{fig:map_rho}, in the plane [density $\rho$] --- [$t_\perp/(t+t_\perp)$], a color map of the Luttinger parameter $K_s$ of the gapless symmetric mode is shown. Except along the line $\rho=1/2$, where there is the gapped RMI, the phase is SF with a continuously varying $K_s\in[1/2,1]$. The map has been obtained using DMRG results. All the limiting cases discussed above are visible here: $K_s=1/2$ along $\rho=0,1/2$ and for all fillings when $t_\perp/t\to\infty$; and $K_s$=1 in the limit $t_\perp/t\to 0$.

\subsection{Disorder effects}

\subsubsection{Weak disorder}

In this section we briefly discuss the pinning of a two-leg ladder boson gas on a random chemical potential. Following Ref.~\onlinecite{Orignac98}, we consider a random chemical potential along both chains. In the continuum,
\begin{equation}
{\cal H}_{\rm dis} = \int dx \left[V_1(x)\rho_1(x) + V_2(x)\rho_2(x)\right].
\end{equation}
We take Gaussian distributions for $V_1$, $V_2$ and assume they are uncorrelated, that is $\overline{V_1(x)V_1(x')}=\overline{V_2(x)V_2(x')}=D\delta(x-x')$ and $\overline{V_1(x)V_2(x')}=0$. RG calculations lead to two very different results whether $t_\perp$ is very weak or very strong. In the case of uncoupled chains ($t_\perp = 0$), the Bose gas is pinned on each chain by the $2\pi \rho$ Fourier component of the random potential as long as $K<3/2$ and for very weak disorder one can estimate the localization length to be \cite{Giamarchi88}:
\begin{equation}
\xi_{\rm loc}^{\rm 1D} = a\left( \frac{\pi v^2}{2Da} \right)^{\frac{1}{3-2K}}.
\end{equation}
The resulting localized phase is called a Bose glass (BG)~\cite{Giamarchi87,Giamarchi88,Fisher89} which can be seen as an incoherent compressible 1D Bose gas. 
Along the line at $t_\perp/t=0$, chains are decoupled and the BG phase consists of two independent strictly 1D BG in which the density is pinned at $2\pi \rho$ so that we use the denomination $BG_{2\pi\rho}$. Its localization length diverges at $K=3/2$, as quantum fluctuations destroy the pinning. 

\begin{figure}
\bc
\includegraphics[width=\columnwidth,clip]{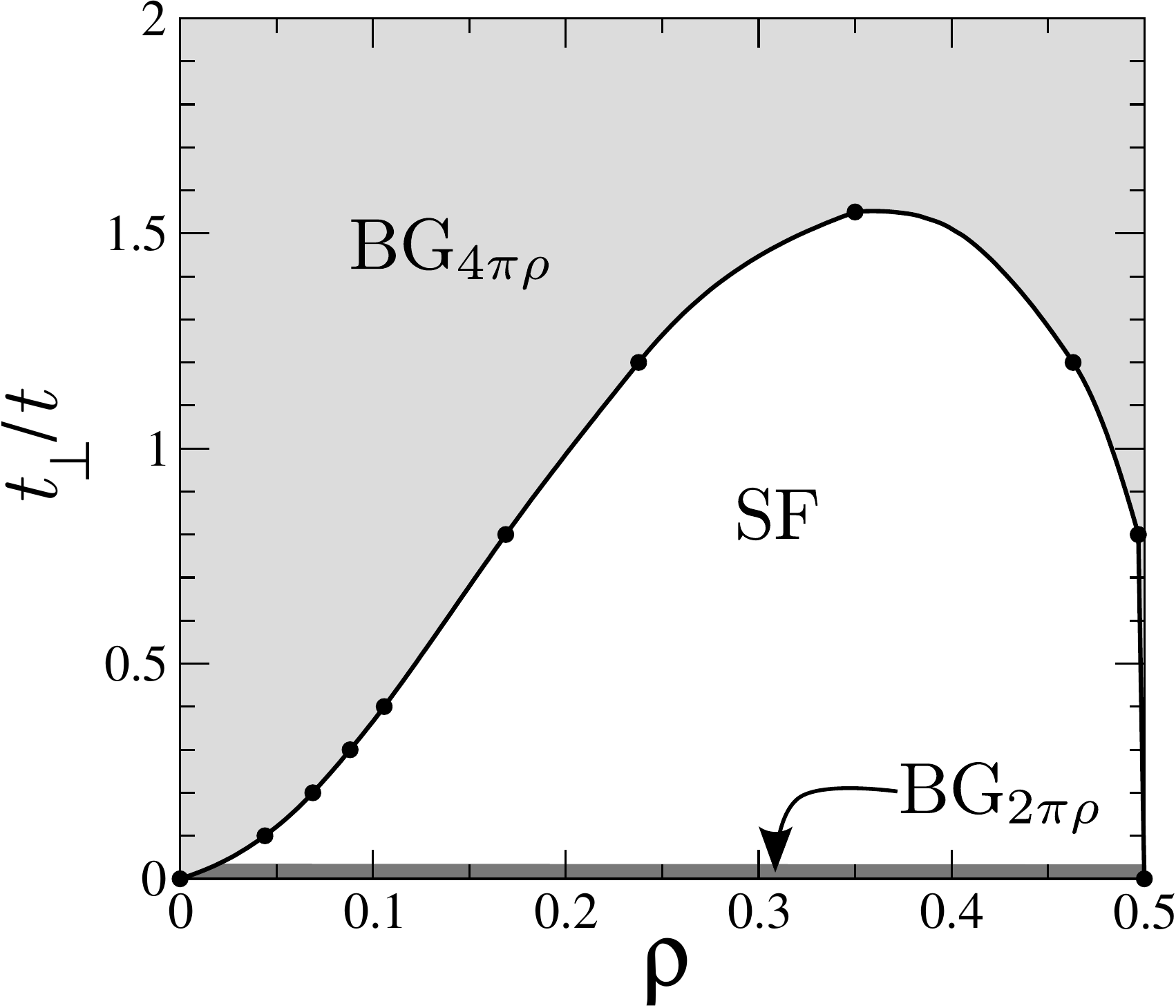}
\ec
\caption{\label{fig:disorder1}(Color online) Phase boundary at $K_s=3/4$ (obtained from DMRG data) separating regions where weak disorder is relevant (for $K_s<3/4$: BG$_{4\pi\rho}$) or irrelevant (for $K_s>3/4$: SF). Along the line at $t_\perp/t=0$, the chains are decoupled and two uncoupled BG at $2\pi\rho$ are expected.}
\end{figure}

On the contrary for very large $t_\perp$, transverse hopping induces long range phase coherence (in the antisymmetric mode), and the superfluid phase is stable against weak disorder provided $K_s>3/4$. On our map of $K_s$ (see Fig.~\ref{fig:map_rho}), it appears that at moderate $t_\perp/t$ (roughly below 1.5) there exists a region of densities for which $K_s$ is indeed above the critical value of $3/4$, making the system insensitive to a random chemical potential. 
In Fig.~\ref{fig:disorder1}, we show the phase boundary $K_s=3/4$ which separates regions where weak disorder $D=0^+$ is relevant (for $K_s<3/4$: BG$_{4\pi\rho}$) or irrelevant (for $K_s>3/4$: SF). In this limit, the transverse coherence length $\xi_a$ introduced in Sec.~\ref{sec:K_SRC} is very short, and one expects a pinning of the Bose gas by the $4\pi\rho$ component of the random potential, hence the denomination $BG_{4\pi\rho}$.

\subsubsection{Finite disorder SF-BG transition}
\begin{figure}
\bc
\includegraphics[width=\columnwidth,clip]{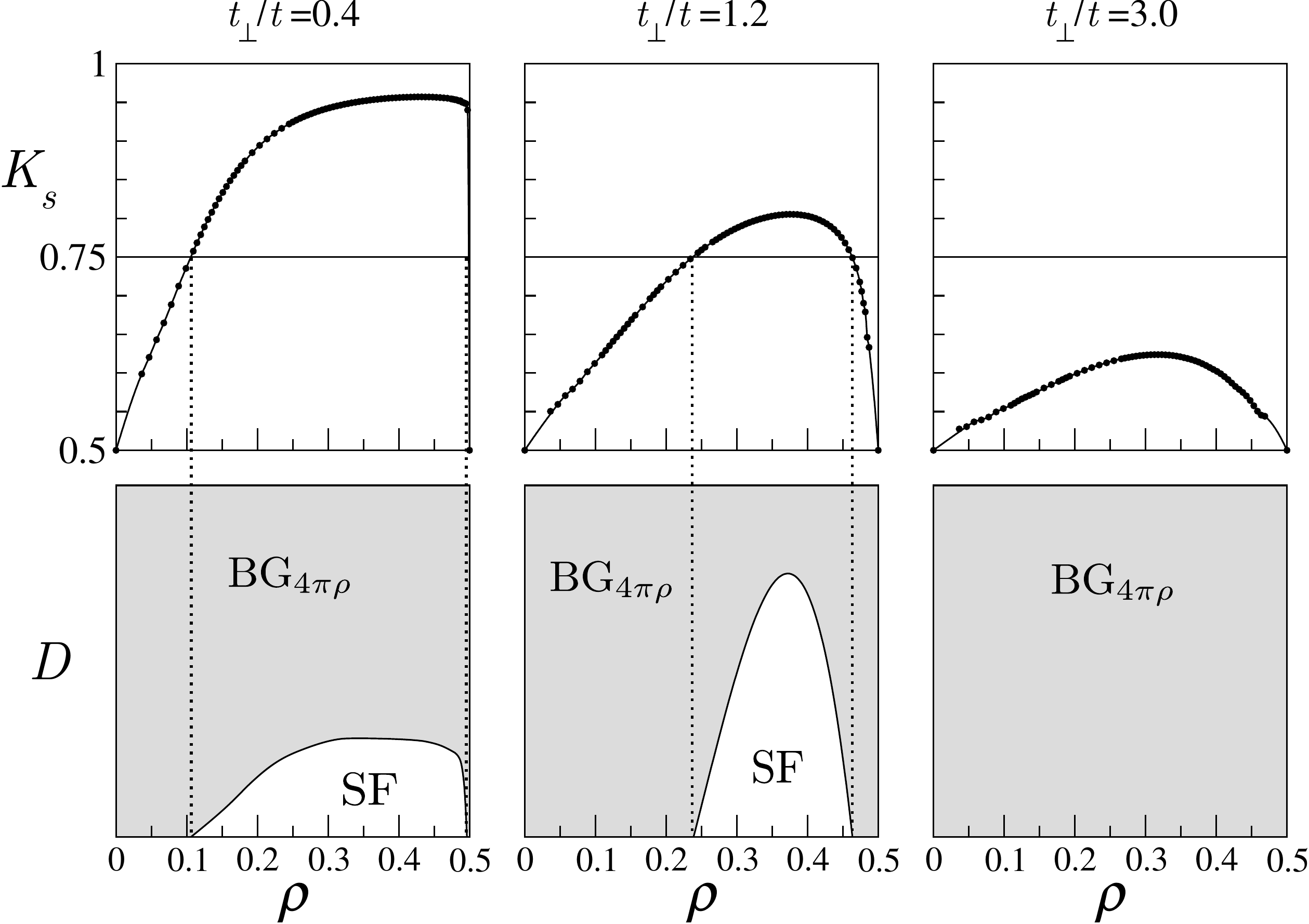}
\ec
\caption{\label{fig:disorder2}(Color online) Three typical situations for 2 coupled chains of hard-core bosons (small, intermediate and strong interchain hopping). On the top line, the Luttinger liquid parameter of the gapless symmetric mode $K_s$ is plotted (DMRG results) versus the density $\rho$. The critical line at $K_s^*=3/4$ is also shown. Bottom: schematic phase diagram in the plane density $\rho$ - disorder $D$ displayed with two different phases, SF and BG at $4\pi\rho$ (see text).}
\end{figure}

One can make use of the criterion $K_s>3/4$ for the stability of the SF phase against weak disorder to plot three representative cases at small, intermediate and strong transverse hopping. For instance at strong $t_\perp/t$, when $K_s$ remains below the critical value of $3/4$ for any filling, one cannot expect a stable SF phase at any finite disorder. This is what is represented in the right panel of Fig.~\ref{fig:disorder2} for $t_\perp/t=3$. More precisely, we expect from the DMRG results for $K_s$ that whenever $t_\perp/t\gtrsim 1.5$, since $K_s<3/4$, weak disorder will immediately lead to localization. This result has to be contrasted with recent numerical studies on disorder bosonic ladders~\cite{Carasquilla11} since we expect strong finite size effects (in particular at low fillings) to be an obstacle to a proper interpretation of numerical results. For intermediate transverse hopping $t_\perp/t\sim 1$, we expect, as shown in the middle panel of Fig.~\ref{fig:disorder2} for $t_\perp=1.2 t$, a stable SF phase for limited fillings but with a maximal critical disorder strength $D{\rm{_c^{max}}}$ not so small since one can naturally expect $D{\rm{_c^{max}}}\sim t_\perp^2$. For the third case at small $t_\perp/t$ (see Fig.~\ref{fig:disorder2} left), we expect a SF phase more extended in term of filling but with a reduced strength, with again $D{\rm{_c^{max}}}\sim t_\perp^2$. The asymmetry of $K_s$ versus $\rho$ will translate into a similar asymmetry for $D_{\rm c}$. At small densities, since $K_s\to 1/2$, we expect the BG phase to extend at weak disorder up to a finite critical concentration $\rho_c$. Close to the gapped RMI phase, while $K_s\to 1/2$ much faster, we still expect a finite window (although much smaller)
with $\rho_c<1/2$ where an intervening BG between RMI and SF phases is present~\cite{Fisher89,Gurarie09}. Again this result has to be contrasted with the numerics obtained in Ref.~\onlinecite{Carasquilla11} at small density and close to half-filling where finite size effects should be quite strong.

In great contrast with the fermionic case where weak disorder would immediately localize the particles, hard-core bosons on a two-leg ladder provide a very simple model where a finite disorder phase transition between SF and BG phases is expected at incommensurate filling. This makes this model quite interesting in the view of large scale numerical studies to precisely investigate the SF-BG universality class, for which to the best of our knowledge, only a few studies exist~\cite{Rapsch99,Roux2008}. In particular, the question of the universal character of the Luttinger parameter right at the critical point for a finite disorder transition appears to be an important question for which two different scenarios have been put forward so far~\cite{Giamarchi87,Altman10}. 

\subsection{Experimental consequences}
\label{sec:exp}

\subsubsection{Coupled Bose liquids in optical lattices}
A necessary requirement to probe our proposed phase diagram with cold atoms is first to obtain
ladder-like structures from an optical lattice. This can be reached by realizing 
a lattice of double-wells which are achieved by superimposing two independent planar optical 
lattices with periodicity $\lambda$ and $\lambda/2$. Such an array of double-wells has been
proposed in Ref.~\onlinecite{Sebby06} and recently realized in Refs.~\onlinecite{Lee07,Chen11} 
with red-detuned lasers (see also Ref.~\onlinecite{Crepin10} for an extensive discussion on the realization of ladder-like potentials with lasers).
A notable result of the present study is the phase diagram depicted in Fig.~\ref{fig:PHDG} where we predict the existence of 
a rung-Mott insulating phase at half filling whatever the transverse hopping parameter.

However, we have shown that the charge gap scales as $\Delta_s\sim \exp(-{\text{a}} t/t_\perp)$ which is exponentially small in the weak transverse hopping regime. Therefore for this phase to be observed, the condition $\Delta_L\ll \Delta_s$ must be satisfied, where $\Delta_L\sim \hbar v/L$ and $L$ is the longitudinal size of the  cigar-shape trap
and $v$ the sound velocity along the chain. Since this size is finite, this puts a constraint on a minimum strength
for the parameter $t_\perp$ in order to observe the rung-Mott insulator. Practically speaking, from Table~\ref{tab:xi}, a relatively large value of $t_\perp/t\sim 1$ is required to make the Mott phase experimentally observable in cold atoms experiments.
Such a Mott insulating phase is characterized by one atom per rung
and should be detectable using  
single-atom resolved fluorescence imaging as recently used for detecting a 2D Mott insulator~\cite{Bakr10,Sherson10}. 

Technically, it seems rather straightforward to achieve the Tonks-Girardeau regime~\cite{Paredes04} for two coupled 1D Bose liquids, as for instance in Ref.~\onlinecite{Chen11}. The system (of total length $L$) being trapped in a harmonic potential, there is a site-dependent chemical potential $\mu_j=V(j-L/2)^2$. Therefore, assuming a zero chemical potential at the trap center $j=L/2$, the Mott state having an energy gap $\Delta_s$ will extend over a finite length 
\be
\ell_{\rm Mott}\approx \sqrt{\frac{\Delta_s}{V}}.
\ee
Using a realistic potential with $V/t=0.003$~\cite{Cheneau11}, the rung-Mott state will be detectable near the center of the trap over $\ell_{\rm Mott}\approx 20$ sites for $t_\perp/t=2$. However, for smaller $t_\perp/t$ the incompressible state will be practically undetectable since $\ell_{\rm Mott}$ decreases quite fast whereas in the same time, the correlation length $\xi$ increases rapidly. 

\subsubsection{Spin ladders in an external magnetic field}
Theoretically predicted some years ago~\cite{Affleck91}, field-induced Bose-Einstein condensation of triplet excitations in a magnetic insulator was first observed experimentally in the spin-1/2 dimer compound TlCuCl$_3$~\cite{Nikuni00,Ruegg03}. As reviewed in Ref.~\onlinecite{Giamarchi08}, a quantitative theoretical description can be achieved through effective theories written in term of hard-core bosons~\cite{Mila98,Totsuka98,Giamarchi99}. 
This is especially
useful for systems which, in zero field, have a singlet-triplet gap in the excitation 
spectrum, such as e.g. spin-1 chains, spin-1/2 ladders, or more complex 3D arrays made of coupled spin-1/2 dimers for instance. While a true Bose condensed state is only possible for D$>1$ systems, a spin ladder in an external field is nevertheless expected to achieve a Luttinger liquid, with SF quasi-long-range order. Recently, an excellent experimental realization for such a Luttinger liquid physics was discovered in the ladder material (C$_5$H$_{12}$N)$_2$CuBr$_4$ (BPCB)~\cite{Klanjsek08,Ruegg08,Thielemann09a,Thielemann09b} for which a direct computation of the Luttinger parameters $K$ and $v_c$ has been performed~\cite{Bouillot11} using various numerical and analytical techniques. BPCB is well described by a spin-$\frac{1}{2}$ antiferromagnetic Heisenberg model on a ladder 
\be
{\cal{H}}=
J\sum_{j,\ell=1,2}{\vec{S}}_{\ell,j}\cdot{\vec{S}}_{\ell,j+1}+J_{\perp}\sum_{j}{\vec{S}}_{1,j}\cdot{\vec{S}}_{2,j}
\label{eq:model_ladder}
\ee
with $\gamma=J/J_\perp$. Nuclear magnetic resonance (NMR) and inelastic neutron scattering (INS) both give $\gamma\sim 0.25$~\cite{Klanjsek08,Thielemann09a}, which puts this material in the class of strong rung couplings. As performed in Sec.~\ref{sec:EffMod} and in the work by Bouillot {\it et al}.~\cite{Bouillot11}, one can write down an effective model with hard-core bosons on a chain with correlated hopping and nearest-neighbor interactions. Here, interactions are repulsive because of the antiferromagnetic exchange that is only slightly shielded by the second-order attractive interactions. Although the perturbative approach developped in Sec.~\ref{sec:K_SRC} is unable to describe quantitatively the evolution of $K$ with the magnetization per rung $m_z$ -- mainly because of the large value of the repulsion in the effective model -- it reproduces some of its features. It correctly gets the limits $K \to 1$ as $m^z \to 0$ and $1$. It also sheds light on the competing effects of correlated hoppings and repulsive nearest-neighbor interactions. Correlated hopping tends to renormalize $K$ upwards, which explains the departure from the values of $K$ obtained from the Bethe ansatz solution of the $XXZ$ chain. Its effects are more pronounced for $m^z<1/2$, since next-nearest neighbor hopping is shielded as the filling of the triplon band approaches $1$. The case of BPCB is interesting in the sense that it lies on the strong rung coupling side with $\gamma \sim 1/4$ where the Luttinger liquid physics in the strong coupling limit can be tested. In the other limit ($\gamma \gtrsim 1$) a Luttinger liquid regime is also expected, as recently observed in the compound DIMPY by Hong and co-workers~\cite{Hong10} with  $\gamma \sim 2$. Nevertheless, as discussed previously for the case of hard-core bosons, we expect a strong anisotropy in the Luttinger parameters, as observed for the velocity in DIMPY~\cite{Hong10}.

One can also mention two other interesting materials: IPA-CuCl$_3$ and BiCu$_2$PO$_6$ which both have a sizeable spin gap $\Delta_s\sim 10$ T for the former~\cite{Masuda06,Garlea07} and $\sim 22$ T for the latter~\cite{Koti07,Tsirlin10}. They both display interesting ladder-like features with some remarkable peculiarities, namely a strong ferromagnetic diagonal coupling in the case of IPA-CuCl$_3$~\cite{Fischer11} and a presumably non-negligible intra-chain frustration in the case of BiCu$_2$PO$_6$. These two effects may lead to a non-trivial field-dependent Luttinger liquid physics. Moreover, the exploration of disorder effects have recently started in these materials~\cite{Bobroff09,Alexander10,Hong10b} where one expects a Bose glass physics to occur when the gap is closed by the external magnetic field. An interesting situation would be to observe a SF-BG transition induced by the external field. Of course a dimensional crossover is also expected at low temperature~\cite{Giamarchi10}.

\section{Conclusions}
\label{sec:conc}

In this paper, we have presented an exhaustive study of the behavior of hard-core bosons on 
a two-leg ladder. The two-leg ladder is a simple minimal geometry that allows for particle exchange, in contrast with strictly one dimensional chains (limited to nearest neighbor hopping).
Therefore, particle statistics should affect the physical behavior of the system.
Using a combination of several analytical and numerical techniques, 
we have indeed shown explictly that the phase diagram of hard-core bosons on a two-leg ladder 
differs significantly from the one of free fermions as 
summarized in Fig. \ref{fig:PHDG}.
The most stringent difference occurs at half-filling and small transverse hopping where hard-core
bosons enter a rung-Mott insulating phase while free fermions are still in a metallic phase.
Outside half-filling, hard-core bosons are in a superfluid phase which is characterized by the Luttinger liquid exponent $K_s$ which varies considerably with the density between $1/2$ and $1$. 
This result is well summarized in Fig. \ref{fig:map_rho} where $K_s$ is plotted as a function of the density and the transverse hopping amplitude. The Luttinger liquid parameter governs all correlation functions but also tells us whether the system might be sensitive to weak disorder or not.
Our quantitative determination of $K_s$ therefore allowed us to draw qualitatively a phase diagram
for hard-core bosons on a two-leg ladder with weak disorder (see Figs \ref{fig:disorder1} and \ref{fig:disorder2}). In great contrast with the fermionic case where weak
disorder would immediately localize the particles, hardcore
bosons on a two-leg ladder provide a very simple model
where a finite disorder phase transition between a superfluid and
Bose glass phases is expected at incommensurate filling. This
makes this model quite interesting in this respect. A more elaborated analysis of the effects
of disorder on the behavior of hard-core bosons on a two-leg ladder will be the subject of a subsequent analysis~\cite{crepin12}.

\acknowledgements
NL would like to acknowledge F. Becca and J. Carrasquilla for useful discussions, and LPT (Toulouse) for hospitality.
\appendix
\section{Gap at half-filling}
\label{app:1}
We outline here the perturbative calculation to second order in $t/t_\perp$ that led to the expression of the size of the plateau at half-filling. We start from the limit of decoupled rungs $t/t_\perp = 0$. The ground state Hamiltonian reads
\begin{equation}
\mathcal{H}_0 = -t_{\perp}\sum_{j}(b^{\dagger}_{1,j}b_{2,j}^{\dagga}+{\rm h.c.}).
\end{equation}
Four states are available on a given rung $j$: an empty state $|0\rangle_j$, two 1-particle state $|1\pm\rangle_j = \left(|j,1\rangle \pm |j,2\rangle\right)/\sqrt{2}$, and a 2-particle state $|2\rangle_j$. At half-filling, $N = L$, the ground state is a tensor product of $L$ symmetric states,  $|L^{(0)}\rangle = |1+,1+,\ldots,1+\rangle$ and the ground state energy is $E_{L}^{(0)} = -t_\perp L$. Now we turn on longitudinal hopping, with $t\ll t_\perp$. The perturbation Hamiltonian is
\begin{equation}
\mathcal{H}_1 = -t\sum_{j,\ell}(b^{\dagger}_{\ell,j}b_{\ell,j+1}^{\dagga}+{\rm h.c.}).
\end{equation}
Acting on $|L^{(0)}\rangle$, $\mathcal{H}_1$  creates $2L$ particle-hole excitations on nearest-neighbor rungs,
\begin{eqnarray}
H_1|L^{(0)}\rangle = -t\sum_{j=1}^L \left[|1+,1+,\ldots,\underset{(j)}{2},0,\ldots,1+\rangle \right. \nn \\
\left. + |1+,1+,\ldots,\underset{(j)}{0},2,\ldots,1+\rangle\right].
\end{eqnarray}
There is no first-order correction to the ground-state energy since $\langle L^{(0)}| H_1|L^{(0)}\rangle =0$. The second-order correction is of the form\cite{Messiah}
\begin{equation}
E_{L}^{(2)} = \langle L^{(0)}|H_1 Q_0 \frac{1}{E_{L}^{(0)} - H_0} Q_0 H_1|L^{(0)}\rangle,
\end{equation}
where $Q_0$ is a projector, that projects out of the ground-state subspace (composed here of one non-degenerate state). This formula has a very intuitive form: $H_1$ creates particle-hole excitations, that are selected by $Q_0$. Then $H_1$ acts a second time on this excited state. A non zero correction is obtained only if this second process takes the system back to the ground-state subspace. Here the excited states have an energy $-t_\perp(L-2)$ and $Q_0[E_{L}^{(0)} - H_0]^{-1}Q_0$ can be replaced by $Q_0/(2 t_\perp)$. The second order correction reads: $E_{L}^{(2)} = - 2L t^2/(2 t_\perp)$, and we find for the  energy of the $L$-particle state,
\begin{equation}
E_{L} = -t_\perp L  - 2L \frac{t^2}{2 t_\perp}.
\end{equation}
We now turn to the states with $L+1$ particles. They are $L$ times degenerate and are best written in a momentum representation,
\begin{equation} 
|(L+1)^{(0)}_k \rangle = \frac{1}{\sqrt{2L}}\sum_{j=1}^L e^{ikj} |1+,1+,\ldots,1+,\underset{(j)}{2},1+,\ldots,1+\rangle.
\end{equation}
Their ground-state energy is $E_{L+1}^{(0)} = -t_\perp(L-1)$. $H_1$ can either move the extra particle around (to next-nearest rungs) or create $2(L-1)$ particle-hole excitations:
\begin{widetext}
\begin{eqnarray}
H_1|(L+1)^{(0)}_k \rangle = &-&\frac{t}{\sqrt{2L}}\sum_{j=1}^L e^{ikj } \left( |1+,1+,\ldots,1+,\underset{(j+1)}{2},1+,\ldots,1+\rangle + |1+,1+,\ldots,1+,\underset{(j-1)}{2},1+,\ldots,1+\rangle\right. \nn \\
&+& \sum_{p\neq j,j-1} \left[|1+,1+,\ldots,\underset{(p)}{2},0,\ldots,\underset{(j)}{2},1+,\ldots,1+\rangle +|1+,1+,\ldots,\underset{(p)}{0},2,\ldots,\underset{(j)}{2},1+,\ldots,1+\rangle\right] \nn \\ &+& \ \left. |1+,1+,\ldots,1+,\underset{(j-2)}{2},\underset{(j-1)}{0},\underset{(j)}{2},1+,\ldots,1+\rangle 
+|1+,1+,\ldots,1+,\underset{(j)}{2},\underset{(j+1)}{0},\underset{(j+2)}{2},1+,\ldots,1+\rangle \right).\nn \\
\end{eqnarray}
\end{widetext}
The first line is the result of moving the extra-particle to neighboring rungs. It brings a first-order correction to the $(L+1)$-particle state energy, $E_{L+1}^{(1)}(k) = -2t\cos(k)$. The second and third lines involve excited states that will bring a correction to second order: 
\begin{equation*}
E_{L+1}^{(2)} = \langle (L+1)^{(0)}_k|H_1 Q_0 \frac{1}{E_{L+1}^{(0)} - H_0} Q_0 H_1|(L+1)^{(0)}_k\rangle,
\end{equation*}
Excited states have here an energy of $-t_\perp(L-3)$ and we can again replace $Q_0[E_{L+1}^{(0)} - H_0]^{-1}Q_0$ by $Q_0/(2t_\perp)$. Acting a second time $H_1$ will shift the energy by a constant (acting on the states on the second line) as it did for the energy of the $L$-particle state. It will also move the extra particle to next-nearest rungs (acting on the states on the third line). The second order correction finally reads $E_{L+1}^{(2)}(k) = -2(L-2)t^2/(2t_\perp) - 2t^2/(2t_\perp) \cos(2k) $, and the corrected energy for the $L+1$-particle state is:
\begin{eqnarray}
E_{L+1}(k) &=& -t_\perp(L-1) -2t\cos(k) \nn \\
&-&2(L-2)\frac{t^2}{2t_\perp} - 2 \frac{t^2}{2t_\perp} \cos(2k).
\end{eqnarray}
Note that we could do the exact same work to compute the energy of the state with $L-1$ particles and would find the same energy. Thus, the size of the plateau at half-filling is given by $\Delta_s = 2[E_{L+1}(0) - E_{L}]$:
\begin{equation}
\Delta_s = 2t_\perp -4t + \frac{2t^2}{t_\perp}.
\end{equation}
\section{Effective model}
\label{app:effmodel}
One can use a somewhat different method to obtain the effective Hamiltonian to second order in perturbation theory for incommensurate fillings~\cite{Mila98,Totsuka98}.  We start from the original model that we cast into, ${\cal{H}}={\cal{H}}_{0} + {\cal{H}}_1$, with:
\bea
{\cal{H}}_1&=&
-t\sum_{j,\ell}\left(b^{\dagger}_{\ell,j}b_{\ell,j+1}^{\dagger}+{\rm h.c.}\right)-(\mu-\mu_c)\sum_{j,\ell}n_{\ell,j}\nonumber\\
{\cal{H}}_0&=& -t_{\perp}\sum_{j}\left(b^{\dagger}_{1,j}b_{2,j}^{\dagger}+{\rm h.c.}\right) - \mu_c \sum_{j,\ell=1,2}n_{\ell,j}\nn \\
\label{eq:model2}
\eea
and $\mu_c = t_\perp$. For the ground state Hamiltonian ${\cal{H}}_0$, states $|1+\rangle_j$ and $|2\rangle_j$ have the same energy $E_0=-2t_\perp$. Let's call $P_0$ the projector on the fundamental subspace, made of $2^L$ states. In this subspace $|1+\rangle_j$ and $|2\rangle_j$ are the only states allowed on a given rung $j$. Let us call $Q_0$ the projector on the complementary subspace. The effective Hamiltonian to second order is:
\begin{equation}
{\cal{H}}_{\text{eff}} = P_0 {\cal{H}}_1 P_0 + P_0 {\cal{H}}_1 Q_0 [E-H_0]^{-1} Q_0{\cal{H}}_1 P_0,
\end{equation}
with $E$ the eigenvalue of degenerate subspace under consideration. Here, given the degenerate ground state and the form of ${\cal{H}}_1$, the {\it virtual} excited state will be a state with an empty rung. Then, in everything that follows, $Q_0[E-H_0]^{-1}Q_0 = Q_0[-2t_\perp L + 2t_\perp(L-1)]^{-1}Q_0= -Q_0/(2t_\perp)$. In the first order contribution the action of the projectors $P_0$ can be enforced by using spin-1/2 operators instead of bosons:
\begin{eqnarray}
 b^\dagger_{j,1} =& b^\dagger_{j,2} =& \frac{1}{\sqrt{2}}\sigma_j^+,\\
b_{j,1} =& b_{j,2}  =& \frac{1}{\sqrt{2}}\sigma_j^-,\\
n_{j,1} =& n_{j,2}  =& \frac{1}{2}\left[\left(\sigma_j^z+\frac{1}{2}\right)+1\right].
\end{eqnarray}
Then, 
\begin{equation}
P_0{\cal{H}}_1P_0 = -t \sum_j \left( \sigma_j^+\sigma_{j+1}^- + H.c. \right) -(\mu -\mu_c) \sum_j \left(\sigma_j^z+\frac{3}{2}\right).
\end{equation}
Now we need to compute $P_0 {\cal{H}}_1 Q_0 {\cal{H}}_1 P_0$. We have seen above in the previous calculation that the only way to create an excited state is to let ${\cal{H}}_1$ empty a singly occupied rung. Let $j$ be such a rung. Then the term of interest in $Q_0 {\cal{H}}_1 P_0$ is of the form 
\begin{equation}
(1-n_{j,1}-n_{j,2})( b^\dagger_{j+1,1} b^\dagga_{j,1}+b^\dagger_{j+1,2} b^\dagga_{j,2})P_0
\end{equation}\\
Now one must act with $P_0 {\cal{H}}_1 Q_0$ on the latter term and bring back the system into the original subspace. As we have already seen, there are two ways of doing so:\\

\noindent {\bf a)} Act with $P_0(b^\dagger_{j,1} b^\dagga_{j+1,1}+b^\dagger_{j,2} b^\dagga_{j+1,2})(n_{j+1,1}+n_{j+1,2}-1)$ and get back the initial state. The corresponding contribution to the second order correction is an attractive nearest-neighbor interaction of the form:
\begin{equation}
-\frac{t^2}{t_\perp}\sum_j\left[\frac{1}{2}-\sigma_j^z\right]\left[\frac{1}{2}-\sigma_{j+1}^z\right]
\end{equation}
where we have used the spin $1/2$ operators to enforce the projection.\\

\noindent {\bf b)} Act with $P_0(b^\dagger_{j,1} b^\dagga_{j-1,1}+b^\dagger_{j,2} b^\dagga_{j-1,2})(n_{j-1,1}+n_{j-1,2}-1)$. This generate a next nearest neighbor hopping term of the form:
\begin{equation}
-\frac{t^2}{2t_\perp} \sum_j \sigma_{j+1}^+ \left[\frac{1}{2}-\sigma_{j}^z \right] \sigma_{j-1}^-.
\end{equation}
A Jordan-Wigner transformation transforms the latter Hamiltonian into one of a band of spinless fermions with next-nearest neighbor hopping and attractive interactions:
\begin{eqnarray}
H_{\text{eff}} &=& -t\sum_j\left[ c^\dagger_j c_{j+1} + H.c. \right] - (\mu-\mu_c)\sum_j (n_j+1) \nn \\
&-&\frac{t^2}{2t_\perp}\sum_j \left[ c^\dagger_{j-1} (1-n_{j}) c_{j+1} + H.c. \right] \nn \\&-& \frac{t^2}{t_\perp}\sum_j (1-n_j)(1-n_{j+1}) 
\end{eqnarray}

\section{Calculation of $K_s$ and $v_c$ in the strong coupling regime}
\label{app:K}

We briefly outline here the calculation that led to the perturbative expressions \eqref{eq:K_SRC} and \eqref{eq:vceff} for $K_s$ and $v_c$. We start from the effective Hamiltonian of equation \eqref{eq:Hefffermions}, derived in section \ref{sec:EffMod}, and extract its Luttinger liquid parameters $\tilde{v}$ and $\tilde{K}$. The low-energy form of the Hamiltonian is \cite{Haldane80}:
\beq
H = \frac{\hbar }{2\pi}\int dx \left[ v_J (\nabla \theta)^2+ v_N (\nabla\Phi)^2 \right],
\eeq

\noindent where the density and phase stiffness $v_N$ and $v_J$ are related to the ground-state energy as \cite{Cazalilla2004}:
\bea
&v_N = &\frac{L}{\hbar \pi} \left(\frac{\partial^2 E_{GS}}{\partial N^2} \right)_{N_0,\varphi=0}, \\
&v_J = &\frac{L\pi}{\hbar }\left(\frac{\partial^2 E_{GS}}{\partial \varphi^2} \right)_{N_0,\varphi=0}.
\eea
Here, $\varphi$ is a twist angle imposed on the boundary conditions as $c^\dagger_{j+L} = e^{i\varphi}c^\dagger_j$ and $N_0$ is the number of particles. Following Cazalilla\cite{Cazalilla04}, the ground-state energy is computed perturbatively to first order in $t/t_\perp$ as $E_{GS} = \langle H_0 \rangle_0 + \langle H_{1} \rangle_0$, with 
\bea
&H_0 =& -t\sum_j\left[ c^\dagger_j c^\dagga_{j+1} + {\rm h.c} \right]\\ 
&H_1 =& -\frac{t^2}{2t_\perp}\sum_j \left[ c^\dagger_{j-1} (1-n_{j}) c^\dagga_{j+1} + {\rm h.c} \right] \nn \\
&&- \frac{t^2}{t_\perp}\sum_j (1-n_j)(1-n_{j+1})
\eea
Using Wick's theorem we find:
\bea
&E_{GS} = &-tL(f_1+f_{-1})\nn \\
&&\hspace{-0.8cm}-\frac{t^2}{2t_\perp}L\left[ (1-f_0)(f_2+f_{-2})+(f_1-f_{-1})^2+2f_0^2\right],\nn \\
\eea
with $f_p = \langle c^\dagger_{j+p}c^\dagga_j\rangle_0$. Note that $f_0 = \tilde{\rho}=N_0/L$ is the density of spinless fermions. Leaving aside the algebra we find, when taking $L\rightarrow \infty$:
\bea
&v_N = &2t\sin(\pi \tilde{\rho}) + \frac{2t^2}{\pi t_\perp}\cos(2\pi \tilde{\rho})-\frac{2t^2}{\pi t_\perp}\nn \\
&&+\frac{2t^2}{ t_\perp}(1-\tilde{\rho})\sin(2\pi \tilde{\rho}), \\
&v_J = &2t\sin(\pi \tilde{\rho}) + \frac{4t^2}{\pi t_\perp}\sin^2(\pi \tilde{\rho})+\frac{2t^2}{ t_\perp}(1-\tilde{\rho})\sin(2\pi \tilde{\rho}).\nn\\
\eea
Finally, we identify the Luttinger parameters $\tilde{v}$ and $\tilde{K}$ by using the relations $\tilde{v}\tilde{K}=v_J$ and $\tilde{u}/\tilde{K}=v_N$ (note the connexion with equations \eqref{eq:KQMC} and \eqref{eq:vc} in the main text) and find:
\bea
\tilde{K} &=& 1+ \frac{2t}{t_\perp}\frac{\sin(\pi \tilde{\rho})}{\pi}, \\
\tilde{v} &=& 2t\sin(\pi \tilde{\rho})\left[ 1+ \frac{2t}{t_\perp}(1-\tilde{\rho})\cos(\pi \tilde{\rho})\right].
\eea
We argue in the main text that for large values of $t_\perp/t$ the correct identification for the original Luttinger parameters is $K_s = \tilde{K}/2$ and $v_c=\tilde{v}$.

\section{Correlation length from data collapse in the $t-V$ model}
\label{app:collapse}

\begin{figure}
\includegraphics[width=\columnwidth,clip]{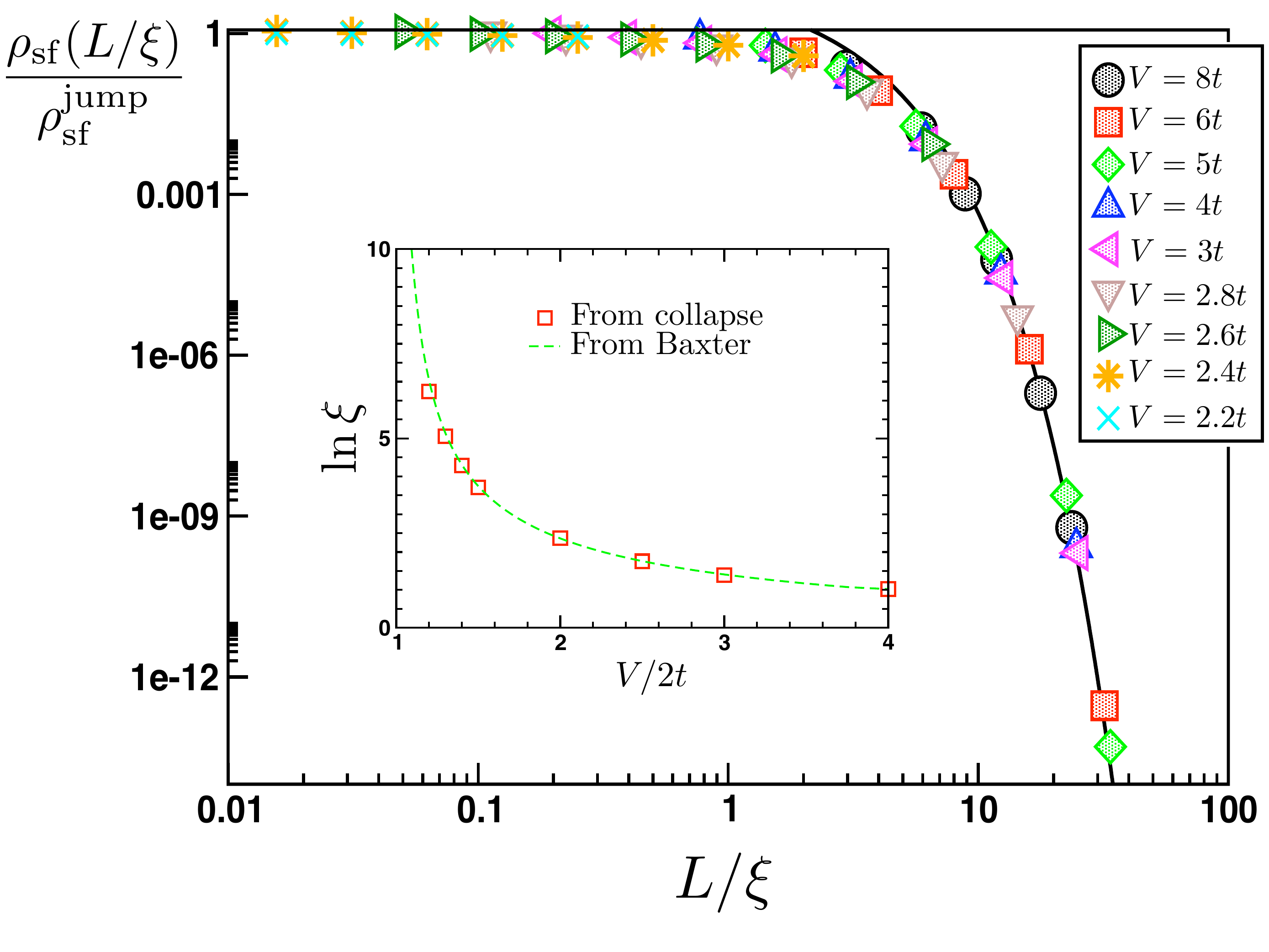}
\caption{\label{fig:XXZ}(Color online) Bethe Ansatz results for the SF density of the integrable $t-V$ chain in the repulsive regime obtained for lengths $L=8,\cdots, 1024$ and various repulsion strengths $V$. Once the data collapse is obtained, the resulting  correlation length $\xi$ is displayed in the inset (red squares) and successfully compared to the exact results from Baxter~\cite{Baxter82} (green dashed curve).}
\end{figure}

In this appendix, we show the usefulness of the ``scaling plot'' technique to obtain the correlation length. As already used in a context of disordered chains~\cite{Laflorencie04} to determine the localization length, it turns out to be a quite efficient method to access (i) the value of the critical coupling, and (ii) the way  the correlation (or localization) length diverges at criticality, even when the numerically accessible system sizes are much smaller than the actual correlation length. To illustrate this technique on a well-controlled example, we briefly look at the hard-core bosonic $t-V$ model (or equivalently fermionic or spin-$1/2$ XXZ) for which one can compute exactly the SF density using the Bethe Ansatz~\cite{Shastry90,Laflorencie01}. This model, governed by the Hamiltonian
\be
{\cal H}_{t-V}=t\sum_i\big[b^{\dagger}_{i}b_{i+1}^{\dagga}+{\rm h.c.}\big]+\sum_iVn_in_{i+1}\,,
\ee
is very well known~\cite{Yang66,Haldane80,Sutherland90}. In particular, there is a SF-Insulator transition where, in the thermodynamic limit, the SF density jumps between $1/4$ and 0~\cite{Shastry90,Laflorencie01} at the Kosterlitz-Thouless transition point $V=2t$. On the insulating side $V>2t$, the correlation length $\xi$, computed exactly by Baxter~\cite{Baxter82}, diverges exponentially. For finite length chains, the SF density vanishes $\sim \exp(-L/\xi)$ while it is finite on the other side of the transition. Using the exact Bethe Ansatz solution of this model with twisted boundary conditions~\cite{Hamer87}, the SF density is computed~\cite{Laflorencie01} and the scaling plot analysis performed above for the ladder system is repeated in Fig.~\ref{fig:XXZ} where one can see the two regimes of Eq.~(\ref{eq:scaling_rhosf}). The correlation length extracted from the collapse is in very good agreement with the exact result~\cite{Baxter82} as shown in the inset of Fig.~\ref{fig:XXZ}.

\end{document}